\documentclass[aps,prb,twocolumn,floatfix]{revtex4}

\usepackage{amssymb}
\usepackage{graphicx}
\usepackage{color}

\def\be{\begin{equation}}
\def\ee{\end{equation}}
\def\bea{\begin{eqnarray}}
\def\eea{\end{eqnarray}}

\begin{document}
\title{Phonon localization in surface-roughness dominated nanowires}
\author{P. Marko\v{s}}
\affiliation
{Department of Experimental Physics, Comenius University in  Bratislava, 842 28 Bratislava, Slovakia}
\author{K. A. Muttalib}
\affiliation
{Department of Physics, University of Florida, Gainesville, FL 32611-8440, USA}
\today
\begin{abstract}
Studies of possible localization of phonons in nanomaterials have gained importance in recent years in the context of thermoelectricity where phonon-localization can reduce thermal conductivity, thereby improving the efficiency of thermoelectric devices. However, despite significant efforts, phonon-localization has not yet been observed experimentally in real materials. Here we propose that surface-roughness dominated nanowires are ideal candidates to observe localization of phonons, and show numerically that the space and time evolution of the energy generated by a heat-pulse injected at a given point shows clear signatures of phonon localization. We suggest that the same configuration might allow experimental observation of localization of phonons. Our results confirm the universality in the surface-roughness dominated regime proposed earlier, which allows us to characterize the strength of disorder by a single parameter combining the width of the wire as well as the mean height of the corrugation and its correlation length.

\end{abstract}

\maketitle
 
\section{Introduction}
Anderson localization \cite{anderson} has been studied most intensely in electronic systems \cite{lee-rama,kramer-mac,markos}, where experiments have clearly observed the metal-insulator transition in three dimensions and absence of true metallic behavior in one and two dimensions as predicted by the scaling theory of localization \cite{abrahams}. The corresponding problem of photon localization has also been studied extensively \cite{john}, although it is only expected to occur in artificially constructed dielectric microstructures \cite{john1,ski}. On the other hand while artificially constructed elastic networks \cite{tigg} do show phonon localization, and numerical studies of finite size properties of inverse participation ratios \cite{garel}  and finite-time scaling \cite{beltukov} clearly indicate the presence of Anderson localization of acoustic phonons in mass-disordered harmonic crystals, the absence of experimental observations of phonon-localization in real materials indicates that either the strength of bulk disorder required is impractical, or that the experimental signature of localization is not clear, or both.

The subject of phonon-localization in real materials, especially in low dimensional nanostructures, has become a topic of current interest because of the role it plays in the context of thermoelectricity \cite{snyder,dubi,taka,hu}. A good thermoelectric device should have a large electrical but a small thermal conductivity, an ``electron-crystal and phonon-glass'' \cite{snyder,slack}. Indeed it was shown in recent experiments \cite{li,hochbaum,lim} that crystalline silicon nanowires with corrugated surfaces can have very small thermal conductivity (reaching the amorphous limit for wires of thickness $d \sim  50$ nm).  It has been argued that such small thermal conductivity can result generically from the presence of localized phonons in nanowires with rough surfaces \cite{ma}. Subsequent numerical simulations \cite{mm} show that in the surface-roughness dominated case, there exists a universal regime where the disorder of the wire can be characterized by a single combination of three different relevant parameters, the width of the wire $d$, the mean height of surface corrugation $h$ and the correlation length $l_c$ of the disorder.  

While the effect of surface disorder on phonon transport has been studied numerically using a variety of techniques \cite{MC1,MC2,MD1,MD2,MD3,MD4,WS1,WS2,WS3,WS4,WS5,WS6}, in the present work we suggest  that it should be possible to experimentally observe the localization of phonons in rough nanowires by systematically studying the frequency and disorder dependence of the propagation of elastic waves when a well-characterized pulse source is injected into the material \cite{majumdar}. By numerically analyzing the time evolution of the energy injected by the source inside surface-roughness dominated nanowires of different disorder, we show that there are several characteristic properties that can be used to identify phonons that have localization lengths much smaller than the length of the wire. The basic idea is very simple; injecting heat-pulses corresponding to various frequency regimes and observing the energy $E(x,t)$ at well-defined positions as a function of time provides a reliable measure of energy localization. We show that for strong disorder, $E(x, t=t_0)$ at a given time $t_0$ clearly shows long-lived resonances excited at positions that vary depending on the frequency, which are hallmarks of existence of localized phonons. At the same time $E_s(t)$ defined as the energy remaining within a small range around the site of injected heat-pulse remains independent of time indicating localization of energy. In addition we show that in this strong disorder regime an appropriately defined mean displacement $r^2(t)$ measuring the diffusion of energy from the injection site also becomes essentially independent of time after a transient period, the saturation value $r_s$being smaller for larger disorder. In contrast, in a weakly disordered regime we observe $E_s(t)$ to decrease with time and $r^2(t)$ to either keep increasing or slowly saturating at a value $r_s$ close to its maximum, indicating that energy is transported easily, although not necessarily in a standard diffusive manner. It should be possible to distinguish between the  two regimes experimentally, by carefully measuring the energy profile across the wire, starting from the site of injection of a well-defined heat-pulse. 
Moreover we confirm that for a given length $L$ of the nanowire, while the surface roughness parameters $h$ and $l_c$ as well as the diameter $d$  of a nanowire determine the strength of surface disorder, a single combination 
\be\label{one}
z\equiv \frac{l_c^{1/2}d^{3/2}}{h}\ee
characterizes the measure of disorder in the surface-roughness dominated regime as proposed in [\onlinecite{mm}]. Thus it should be possible to study the effects of such disorder systematically by preparing nanowires with similar $z$-values. We suggest that the wires produced by Electroless Etching (ELE) in the experiments of Hochbaum et al \cite{hochbaum} showing amorphous-like thermal conductivity should be good candidates to observe phonon-localization.

\section{Observing localized phonons}

There are two possible reasons why localization of phonons has not yet been observed in real systems although localization of electrons has clearly been observed experimentally. We discuss the reasons and propose alternative options in the following subsections. 

\subsection{Bulk vs surface disorder}

Even in electronic systems, driving a metal to an insulator by increasing the bulk disorder by, e.g., increasing the density of impurities is not easy; the resistivity often saturates when the mean free path becomes comparable to the lattice spacing. The original solution was to dope the material, where it was possible to observe metal-insulator transition by systematically doping phosphorous in silicon \cite{rosenbaum}.  For phonons, obtaining localized states in a silicon wire would presumably require replacing some of the silicon atoms with  `defect' atoms. However as numerical simulations by Monthus and Garel \cite{garel} show, adding defect masses twice as heavy had only a minor effect and kept the system in the weakly disordered regime. Indeed, the simulations required defect masses almost twenty times as heavy in order to observe a transition to localization in three dimensions. This would mean that random substitution of even the heaviest atoms available will not be sufficient to localize phonons in a silicon wire. 
 
An alternative possibility is to consider surface disorder in reduced dimensions. This will not allow us to explore the critical nature of phonon localization transition since all states are expected to remain localized in reduced dimensions. We therefore do not attempt to study the localization transition; instead our focus is on the existence and observation of localization of acoustic phonons in surface-disorder dominated nanowires.  This is also more relevant in the context of thermoelectricity, where a surface disorder will typically affect the thermal conductivity of a nanowire more than the corresponding electrical conductivity and is therefore more suitable as a thermoelectric device. It is already known \cite{hochbaum} that thermal conductivity of surface disordered (but otherwise crystalline) silicon nanowires of diameters $d \le 115$ nm (prepared by ELE) can have very low thermal conductivity, which can reach the limiting amorphous limit for  $d\sim 50$ nm. This was attributed to the presence of localized phonons \cite{mm}. If correct, it should be easier to achieve localization of phonons in surface disordered nanowires.

\begin{figure}[t]
\begin{center}
\includegraphics[width=0.42\textwidth]{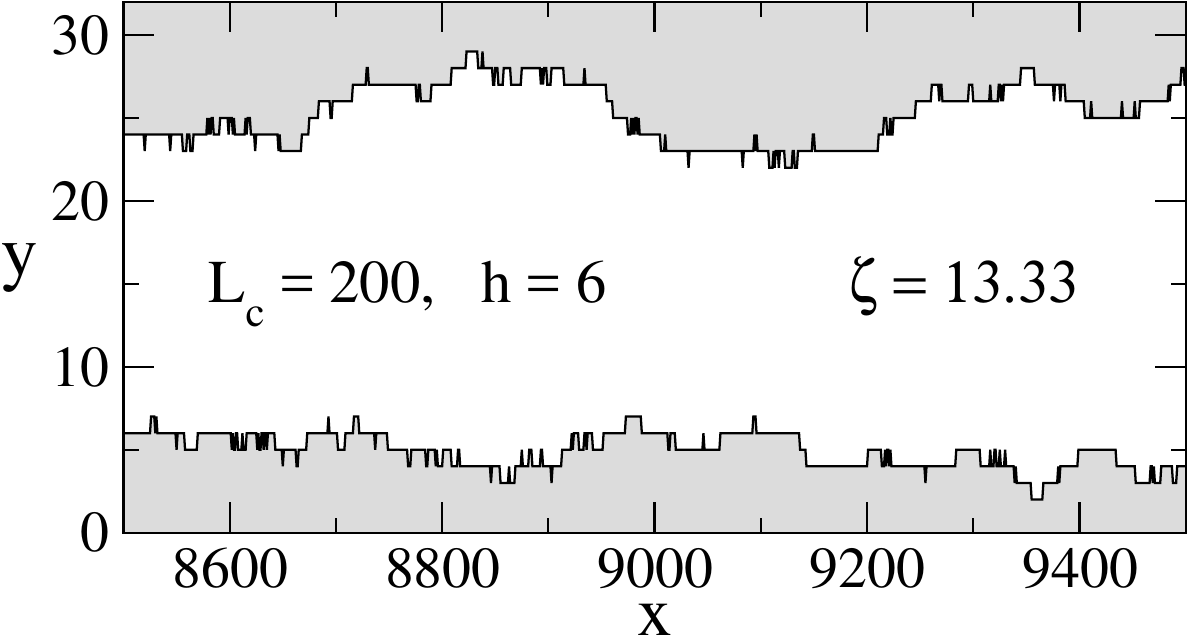}\\[5mm]
\end{center}
\caption{%
	(Color online)
Typical sample under study. The length of the system in the $x$ direction is  $L= 16000$ (only small part of the sample is shown), the width  $d=32$, correlation length $l_c=200$ and disorder $h=6$, corresponding to $z = 427$. We consider absorbing boundary conditions at ends of the wire.
}
\label{samplewire}
\end{figure}

One problem with surface disorder is that it requires several parameters to characterize its strength. Consider Figure \ref{samplewire}, which is a typical sample we use for our numerical simulations.
First of all in addition to the length $L$ of the wire, the diameter $d$ appears explicitly in characterizing the effective disorder since obviously the same surface disorder would be more effective in narrower wires than in wider wires. In addition, the strength of disorder is not only given by the mean height of the corrugation $h$, but also by the correlation length $l_c$. All these parameters are separately used in e.g. characterizing the ELE silicon nanowires used in the experiments of Hochbaum et al \cite{hochbaum}. In order to do a systematic study of localization of phonons, this is clearly too large a parameter space to explore experimentally. However, recently it has been shown by numerical simulations \cite{mm} that a single parameter $z\equiv l_c^{1/2}d^{3/2}/h$, as defined in Eq. (\ref{one}), characterizes the effective disorder of the wire in a regime where surface disorder dominates over any bulk disorder and transport is dominated by diffusive phonons. It is not clear if this characterization remains valid in the strongly disordered systems dominated by the presence of localized phonons. If true, this will clearly  simplify the search for phonon localization enormously.

\subsection{Heat-pulse and energy evolution}

The second complication with phonons is that while for electronic systems the existence of a Fermi surface makes sure that transport is dominated by electrons near the Fermi energy, heat transport involves a sum over a band of phonon frequencies, including very low frequencies that are hard to localize. If, e.g., states beyond a certain frequency are localized, all other states below that frequency remain either diffusive or ballistic and  contribute to the transport. Therefore the signature of the presence of these localized phonons are not obvious from transport measurements, except for a reduction in thermal conductivity. 

Majumdar \cite{majumdar} has suggested that one can inject a heat pulse at some given point on the wire and observe the time evolution by observing the effect at various distances as a function of time. Presumably the effect of the source pulse will reach different distances at different times depending on if it encounters a localized phonon or not. Here we propose that in such an experiment, it should be possible to study the
energy evolution $E(x,t)$ (kinetic and potential) accumulated  at time $t$ at a position $0<x<L$, which includes the total energy of all atoms across the width $d$ of the wire. The total energy  in the system is  
\be E(t) = \sum_x E(x,t).
\label{Eoft}
\ee
For more detailed analysis, it is also useful to consider the energy in a given region (for instance, in the neighborhood of the source), 
\be E_s(t) = \sum_{|x-x_s|<\Delta} E(x,t),~~~~~\Delta \sim 100 a
\label{Esoft}
\ee
where $a$ is the lattice spacing. Clearly for a localized phonon, this quantity should remain almost independent of time. 
As for transport properties, a suitable parameter is the mean displacement, $r^2(t)$, which measures the diffusion of the energy from the source,
\be
r^2(t) = \frac{1}{E(t)}\sum_x \frac{(x-x_s)^2}{12L^2}E(x,t).
\label{r2oft}
\ee
Note that $r^2(t)$ is normalized by factor of $12$ so that  the value $r^2=1$ corresponds to the energy homogeneously distributed along the sample. Thus a localized phonon would lead to a saturation value of $r^2\ll 1$ after a transient time, the saturation value being smaller for larger disorder. 

Of course, the above quantities will depend on the specific realization of disorder. For a more accurate analysis, it would be useful to  repeat the
simulations for an statistical ensemble of disordered samples and consider the mean values, which is beyond the scope of our current investigation. Nevertheless, we emphasize that while such ensemble averaging would produce a more smooth time-dependence of the above quantities, the qualitative signatures of localized phonons as described above in the overall time-dependence are not sensitive to the sample to sample fluctuations.

\section{Model and numerical simulations}

Typical structure for a nanowire with surface disorder is shown in Fig. \ref{samplewire}. Atoms with mass $M_0 = 1$ occupy a rectangular square lattice. The distance between  nearest neighbor 
atoms $a=1$ defines the unit of length. The spring constant $k=1$ measures the harmonic force between neighboring atoms.
For a square lattice, allowed frequencies fill the frequency band $0<\omega<2\sqrt{2}$. The size of the sample is defined by its width $d$ and length $L$.
The surface disorder is modeled by correlated disorder of mean corrugation height $h$ and correlation length $l_c$.
Beside the four length parameters  $d$, $L$, $l_c$ and $h$, the propagation of phonons depend on their frequency $\omega$ or period $T=2\pi/\omega$.
In our units, the speed of long wavelength  phonons $c = \sqrt{k/M_0} \equiv 1$ so that the value of period $T$
equals to the wavelength, $\lambda = cT$.

The sample is excited by a time dependent force acting on atoms in one column, usually at the center of the sample $x_s=L/2$:
\be 
s(t) = \exp\left[-\frac{(t-t_0)^2}{2\sigma^2}\right]\cos 2\pi t/T
\label{eq:source}
\ee
with $t_0=1500$ and $\sigma=500$. The frequency of the source, $\omega=2\pi/T$ is chosen from the
acoustic band, $0<\omega<2\sqrt{2}$. 

The energy, given by the external force, propagates through the sample. 
In our model the energy is transmitted by scalar acoustic waves propagating  on the two-dimensional square lattice. 
We solve numerically the wave equation  
\bea\label{wave:eq}
\frac{\partial^2 u(x,y)}{\partial t^2} &= &u(x+a,y)+u(x-a,y)\cr
&+&u(x,y+a)+u(x,y-a) - 4u(x,y)
\eea
\textcolor{black}{For each $(x,y)$, Eq. \ref{wave:eq} is an equation of motion of atom interacting with its  four nearest neighbors.
Equation \ref{wave:eq} is formally the finite difference approximation of continuous wave equation with Laplacian $\Delta u(x,y)$  substituted  by expression on the r.h.s. of (\ref{wave:eq}).  
We apply  
explicit numerical algorithm described in [\onlinecite{nr}] with time step $\delta t=T/60$
and  set $u\equiv 0$ along the disordered  horizontal boundaries.} Absorbing boundary conditions \cite{abs-bc} are implemented  at the left ($x=0$) and right ($x=L$) boundaries.

\textcolor{black}{%
Neglecting  phonon-phonon interaction in our model is justified by 
	experimental data for thermal conductivity $\kappa$ of surface-roughness dominated  silicon nanowires \cite{li,hochbaum,lim}, e.g. synthesized by electroless etching (ELE).
Phonon-phonon interactions give rise to a $1/T$ temperature dependence of $\kappa$ at high temperatures \cite{am} in contrast to the power-law  behavior at low $T$, leading to a peak in $\kappa(T)$ at some intermediate temperature. For bulk silicon this peak occurs at around $25$ K, but for ELE nanowires of thickness less than $115$ nm this peak is not observed up to $300$ K, showing the dominance of surface scattering even at high temperatures.   
}

Typical time of simulation ${\cal T} \sim  3\times 10^5$. 
After the source is applied, we  calculate numerically the energy 
$E(x,t)$ (kinetic and potential) accumulated  at time $t$ in the  `column' $x$. Note that this is the energy of all atoms in the column $x$. 
In what follows, we will present the time  and space distribution for the energy 
${\cal E}(x,t) = 100\times E(x,t)/E_{\rm max}$  normalized to maximal energy observed throughout the simulation,
$E_{\rm max} = {\rm max}_{x,t}\{E(x,t)\}$.

\begin{figure}[t]
\begin{center}
\includegraphics[width=0.48\textwidth]{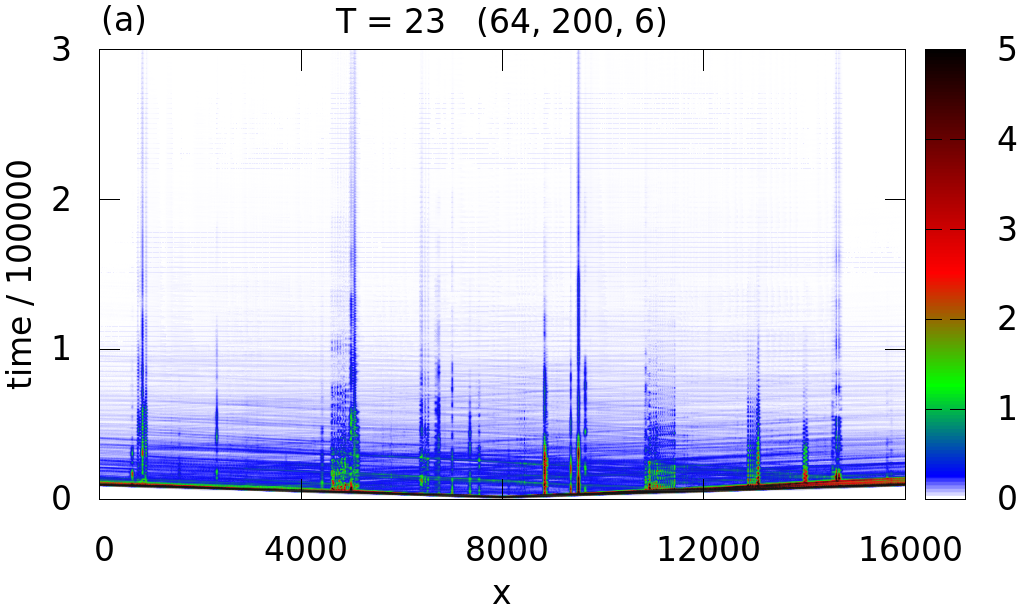}

~~\\
\includegraphics[width=0.23\textwidth]{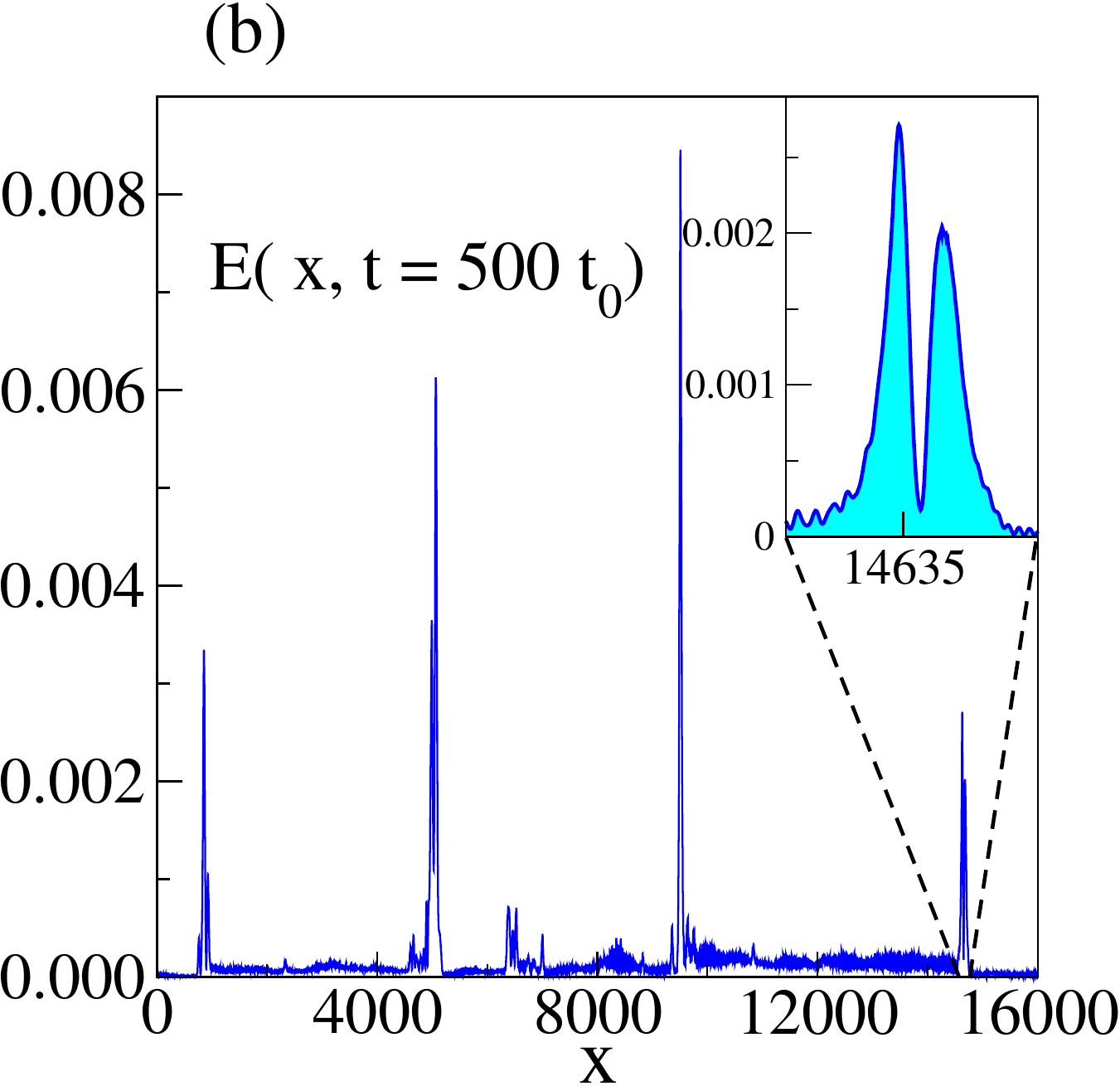}
~~
\includegraphics[width=0.21\textwidth]{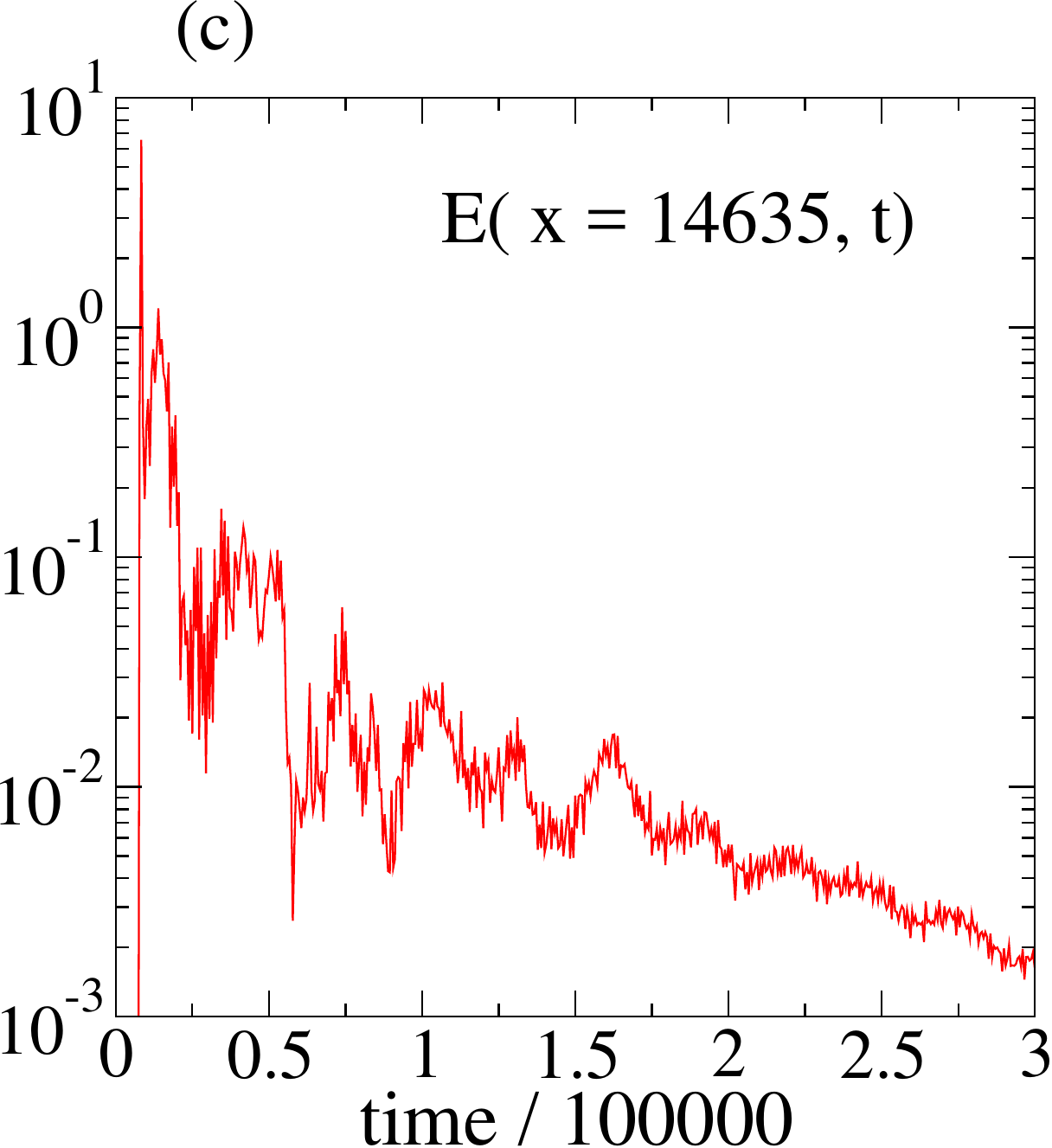}

~~\\
\includegraphics[width=0.48\textwidth]{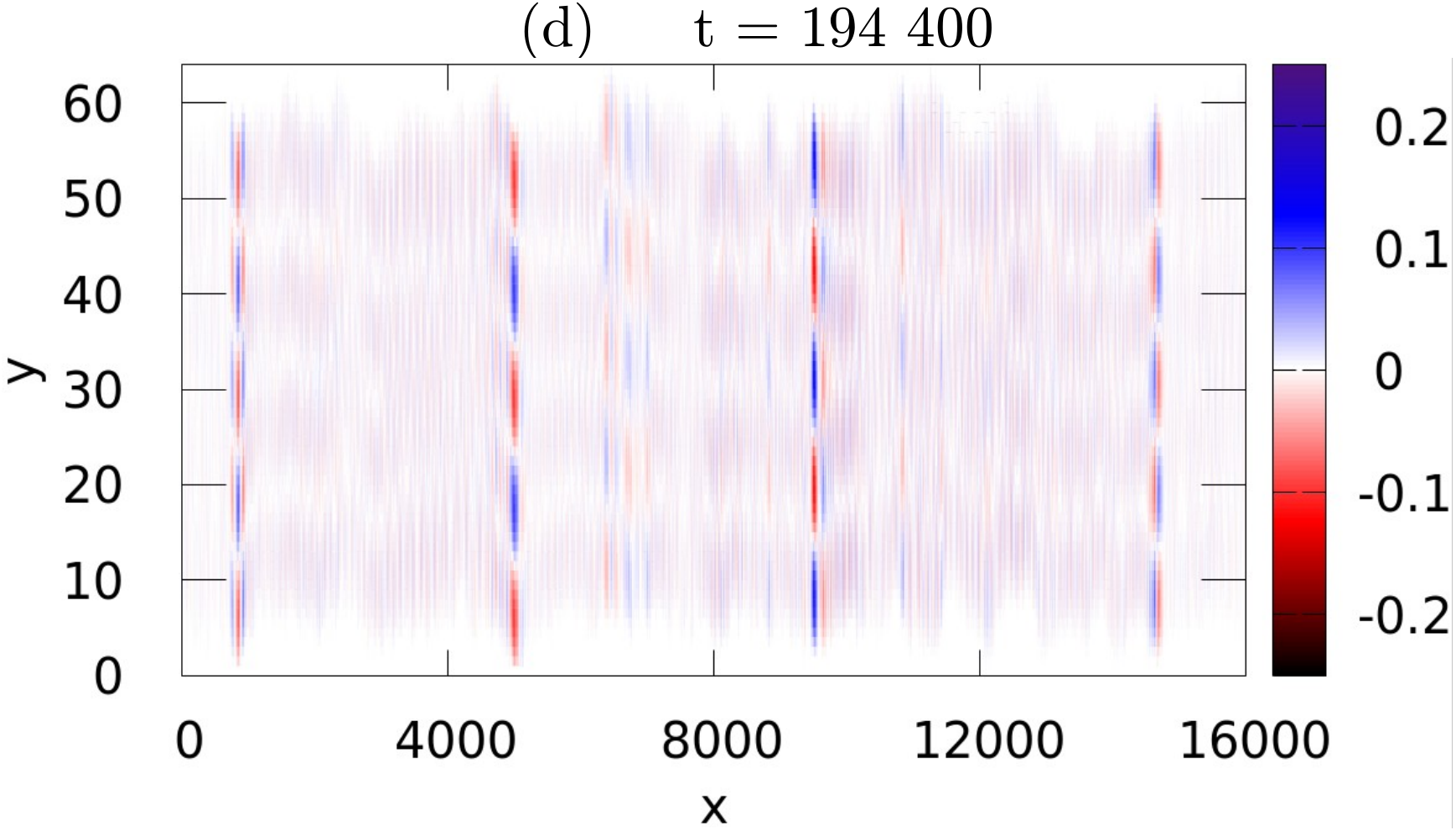}
\end{center}
\caption{%
(Color online)
(a) Space and time evolution  of normalized energy ${\cal E}(x,t)$ in the weakly disordered sample of size
$64\times 16000$. 
Period of the source $T = 23$.
Three length parameters, $(d,l_c,h) = (64, 200, 6)$ are given in the title, 
corresponding to $z = 1206$. 
A set of 
resonances excited by the external source is clearly visible, corresponding to localized phonons.
(b)
The space distribution of the energy at time $t=500t_0$ ($t_0 = 555.55$). Inset shows detail of the resonance at $x=14635$.  (c)  Time evolution of the energy at 
	the position of resonance at $x=14635$. (d) Snapshot of $u(x,y)$ at time $t= 194~400$ offers another method to identify resonances showed in panels (a) and (b). \textcolor{black}{Please note different scale on the horizontal and vertical axes.}
}
\label{amp64-1}
\end{figure}

 \section{Results}

The first question we address  is whether or not localized phonon states exist in our system that can be excited by the injection of a localized heat-pulse. The second question we study is whether the existence of localized phonon states imply localization of energy around the region where the heat-pulse is injected. As we show below, the answer to both is positive.

\subsection{Existence of localized phonon states}

Fig.  \ref{amp64-1}  shows the evolution of the energy in time and space along the sample for a period of the source $T=23$. A small number of long-lived resonances are clearly visible. The energy propagates from the center of the lattice 
(position of the source in this case) and excites resonances which correspond to localized phonons.
The lifetime of resonances can be very long; some are excited at the beginning of the simulation within time  $\sim 10^4$, and  survive the entire time of the simulation ${\cal T}\sim 10^5$.  As shown in Fig \ref{amp64-1}(c), the energy localized at the resonance oscillates in time being transferred to other resonances. This is confirmed by detailed analysis of time evolution (not shown). Finally, Fig. \ref{amp64-1}(d)  presents a snapshot of the energy distribution in the sample for $t = 194 400$. Four resonances, identified also in panels (a) and (b), are clearly visible.

\begin{figure}[b]
\begin{center}
\includegraphics[width=0.48\textwidth]{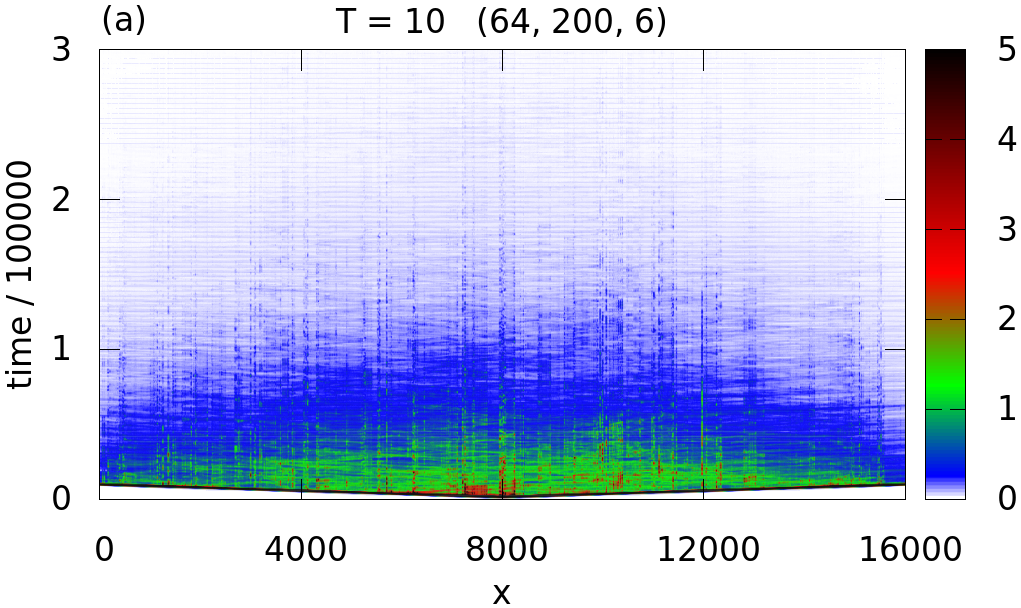}

~~\\
\includegraphics[width=0.48\textwidth]{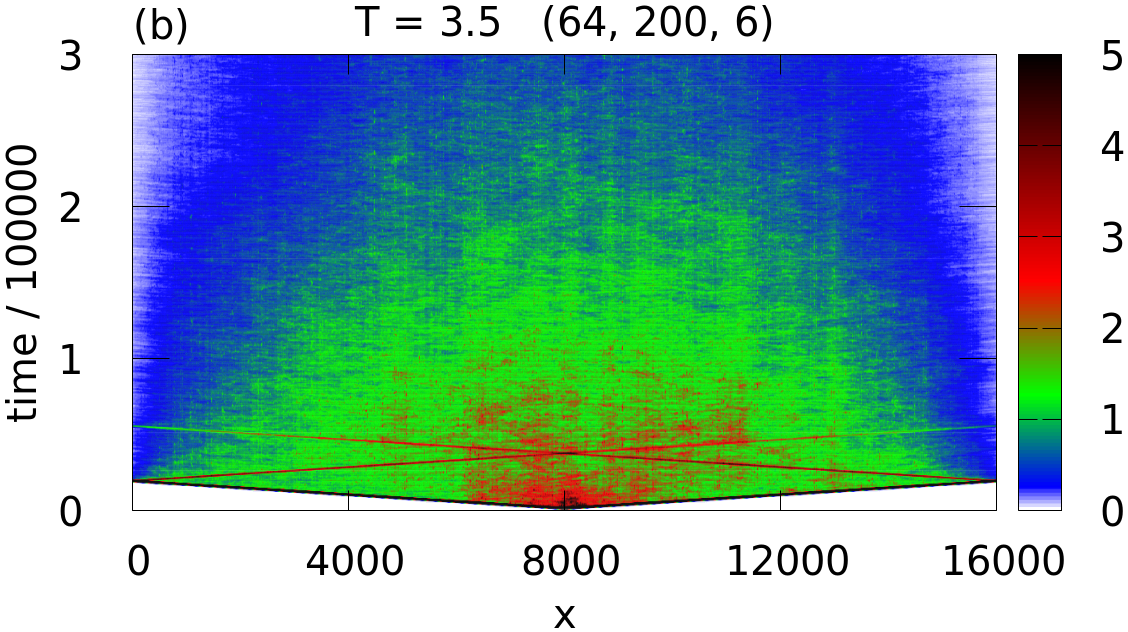}
\end{center}
\caption{%
(Color online)
Space and time evolution of normalized energy ${\cal E}(x,t)$ for the same sample as in Fig. \ref{amp64-1} but with smaller periods of the source, $T=10$ and $T=3.5$.
Note that resonances  exist at all frequencies of the source and that their number increases when period $T$ of the source  decreases. 
Also, resonances lie closer to each other and their lifetimes are shorter.
}
\label{amp64-2}
\end{figure}

We verified (data not shown) that the positions of resonances do not  depend on the position of the source. This means that they could really be associated with localized eigenstates of the disordered structure. On the other hand,
changing the  frequency of the source results in excitation of different resonances localized at other positions of the sample.  
As predicted theoretically \cite{mm}, the number of resonances increases when the frequency of the source increases. 
Since their mutual distance decreases, their lifetime, defined by overlap of eigenstates,  is shorter. As an example,
Fig. \ref{amp64-2} shows the time evolution of the energy for the same sample but at two higher frequencies, $T=10$ and $T=3.5$. Note that the shorter lifetime does not necessarily indicate a wider resonance, but just that two nearby resonances has larger probability to overlap. Clearly, this depends on the particular realization of disorder and an averaging over ensembles with different realizations of disorder will be needed to estimate the range of $T$ where one can expect localization.

\begin{figure}[t]
\begin{center}
\includegraphics[width=0.48\textwidth]{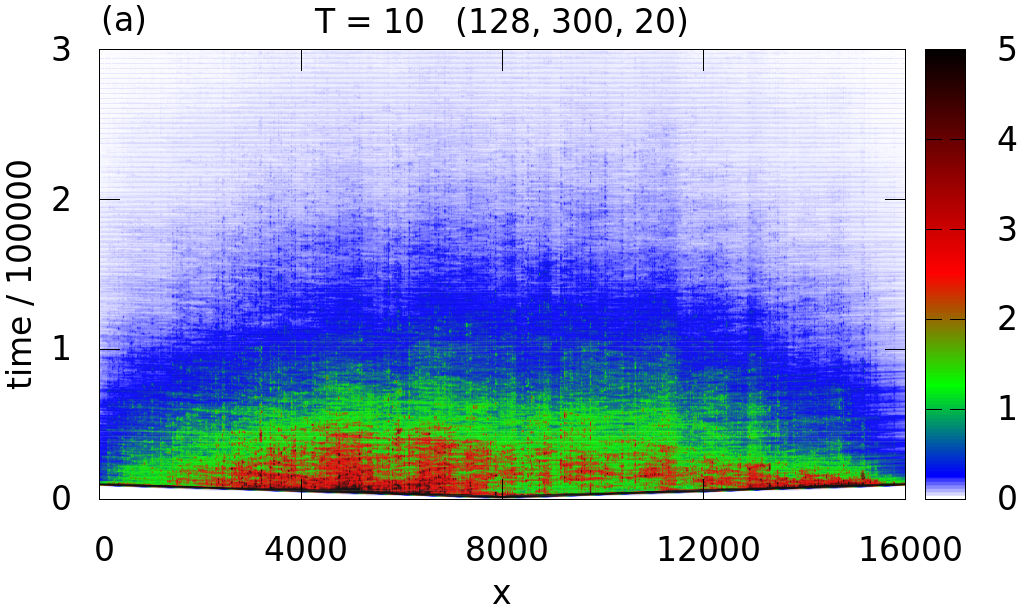}

~~\\
\includegraphics[width=0.48\textwidth]{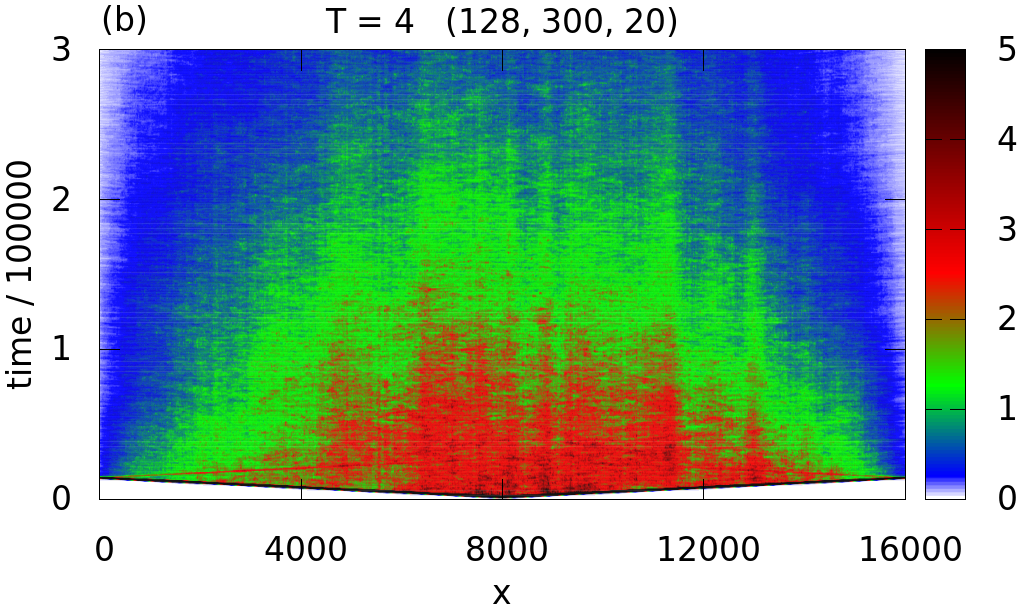}
\end{center}
\caption{%
(Color online)
Space and time evolution of normalized energy ${\cal E}(x,t)$ in sample 
 $128\times 16000$ with   $l_c=300$ and $h=20$  ($z= 1253$)
for periods (a) $T=10$ and (b) $T=4$. 
Note  the similarity of panels for the period $T=10$ with that shown  in Fig.  \ref{amp64-2}.
}
\label{amp128-1}
\end{figure}

We found that localized resonances are not necessarily limited to small widths comparable to the mean corrugation height $h$.  As an example, we show in Fig. \ref{amp128-1} resonances observed in a system with $d=128$ and $h=20$. As we show later, localization depends on a combination $z$ of three parameters of the wire $d$, $h$ and $l_c$ as well as the frequency $\omega$ of the source.

\medskip

\begin{figure}[b]
\begin{minipage}{0.25\textwidth}
\includegraphics[width=0.95\textwidth]{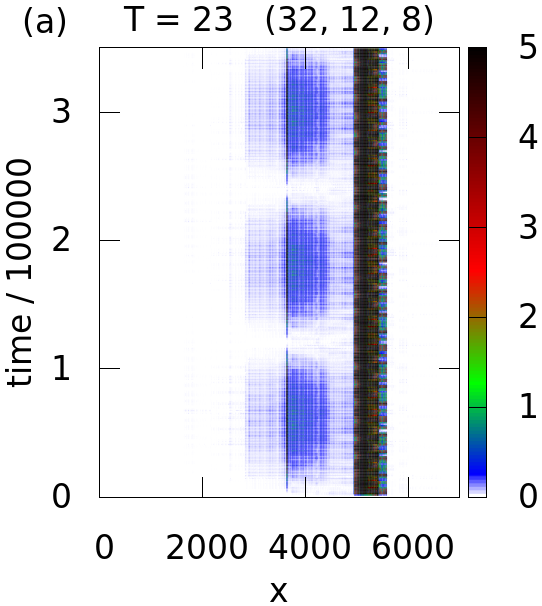}
\end{minipage}
\begin{minipage}{0.21\textwidth}
\includegraphics[width=0.95\textwidth]{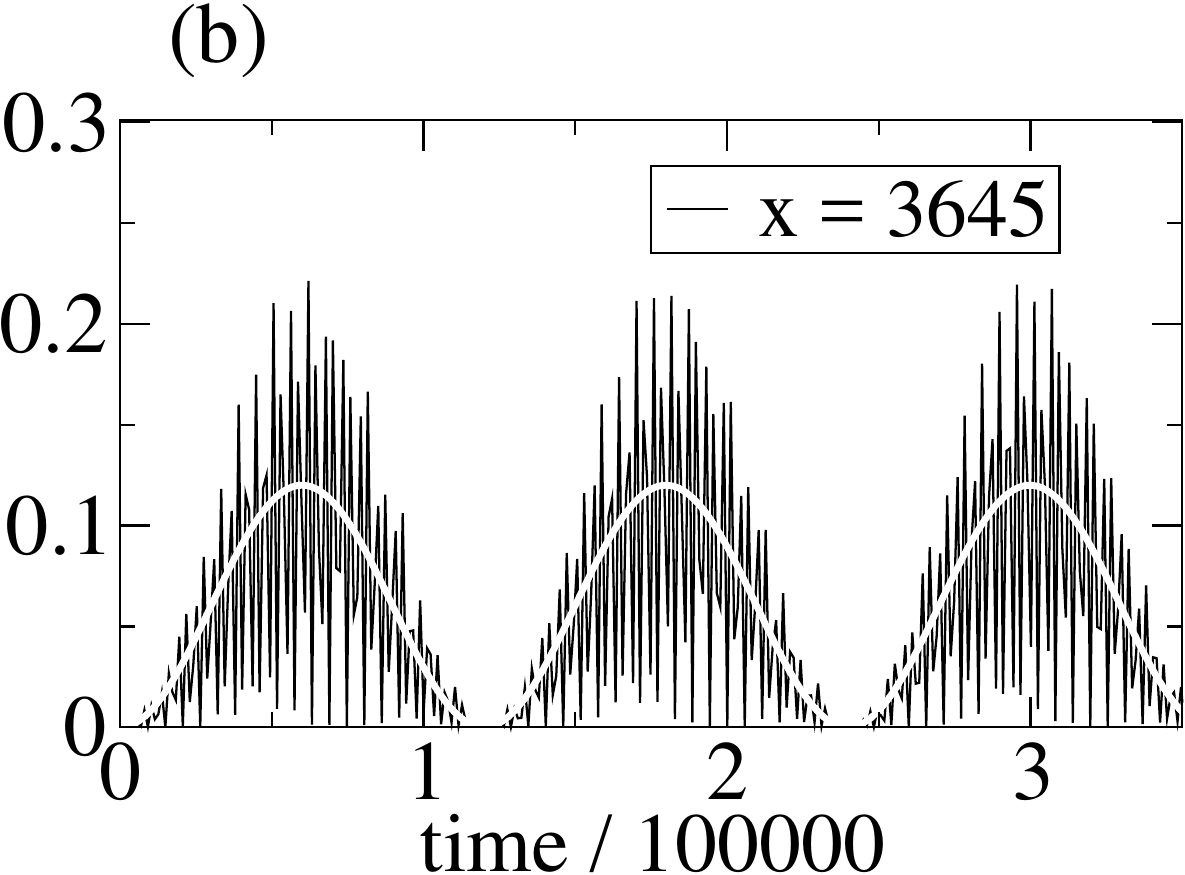}

~~\\
\includegraphics[width=0.98\textwidth]{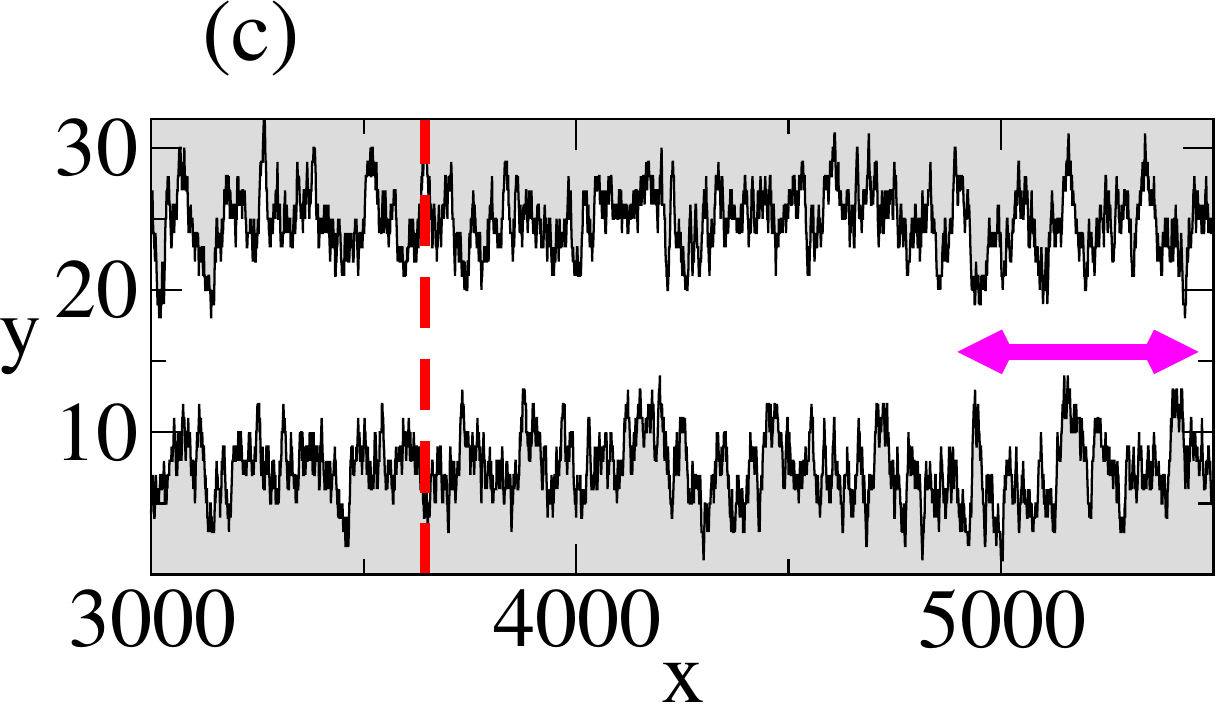}
\end{minipage}
\caption{%
(Color online)
(a)  Localized phonon observed in a narrow and strongly disordered sample. The 
size of the sample is $32\times 10240$, disorder $h=8$ and correlation length $l_c=12$ ($z = 78.38$).
Source is located at $x=5120$ and
period of the source is $T = 23$.  Small portion of energy  tunnels from bottleneck region and excites periodically the resonance localized 
 at $x\approx 3645$. 
(b)  The time evolution at energy $E(x=3645,t)$. White solid line is given by  Eq. (\ref{feynman}) with $\omega -\omega_R = 5.24\times 10^{-5}$.
 (c)   The detail of the structure. Here, vertical  dashed line shows
the position of the localized state and magenta arrow indicates the region where the injected energy is trapped.
}
\label{d32-1}
\end{figure}

For small $d$ and large disorder $h$, it might happen that the width of the wire at some part of the sample becomes 
smaller than the wavelength $\lambda$ of propagating phonons. One example is shown in Fig. \ref{d32-1}, where energy of the source is trapped
in a narrow region in the center of the sample.
This is easy to understand: the phonon 
can propagate through the wire with local width $\tilde{d}(x)$ only if the phonon frequency ${\omega} > 2\pi/\tilde{d}$.
Indeed,  
phonons with higher frequency  easily propagate through the bottleneck region (data not shown).

Localization of the injected energy in a small region shown in Fig. \ref{d32-1}(a)  enables us to demonstrate  the existence of 
localized phonons. The energy trapped in the  narrow region around the source location
tunnels to the left and excites the resonance localized at $x=3645$ (position of the resonance is shown in Fig. \ref{d32-1}(c)). The resonance excites and de-excites with the period $\sim 10^{5}$ (Fig. \ref{d32-1}(b)) following the textbook formula for double-well potential\cite{Fe} 
\be\label{feynman}
E(t)\sim \sin^2\left[ \displaystyle{\frac{\omega-\omega_R}{2}}t \right],
\ee
where $\omega_R$ is the eigenfrequency of the resonance and $\omega = 2\pi/T$. The effect is observable only for frequencies close to $\omega_R$. 
Note that the position of the resonance does not exhibit any special fluctuation of the
surface disorder (in contrast to the position of the ``bottlenecks'').

\subsection{Localization of energy and universality}

In the previous subsection we showed that localized phonon states do exist in wires with large surface disorder. Now we want to show that the existence
of these localized states imply localization of energy injected into the sample. 
To measure  the effect of localization, we calculate the time evolution of the energy $E_s(t)$ defined in (\ref{Esoft}), the total energy $E(t)$ around the region of the source, as well as the mean displacement of energy $r^2(t)$ away from the source.
Time independence of these functions would imply that the energy is not able to propagate away from a given region, which we expect to be a reliable measure of the localization of energy.

\begin{figure}[h]
\begin{center}
\includegraphics[width=0.22\textwidth]{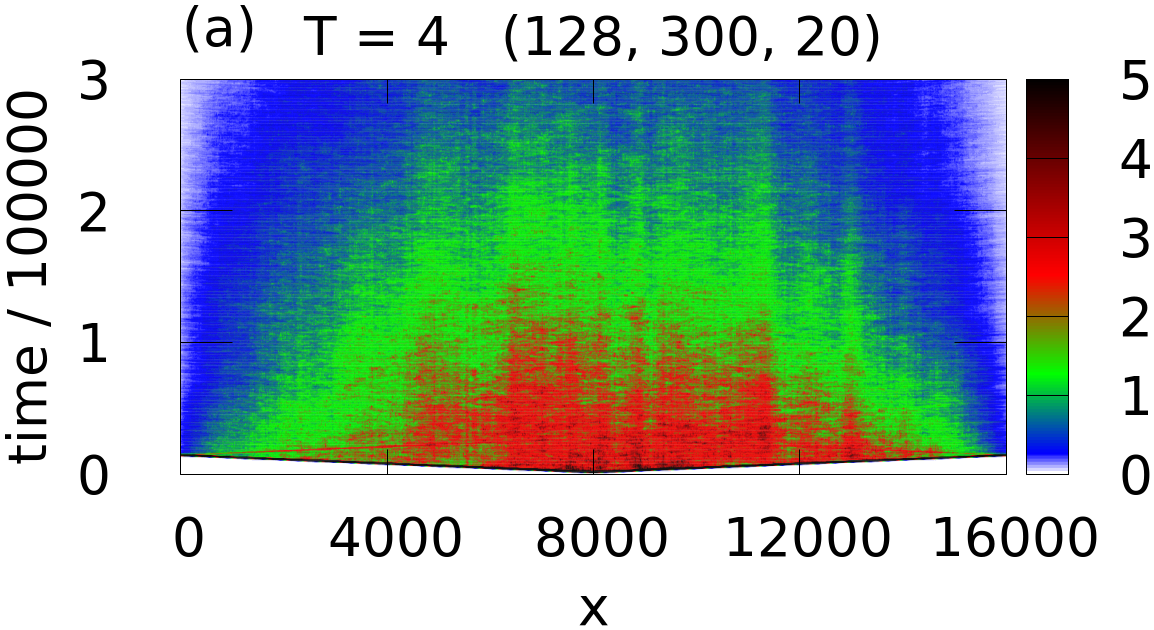}
\includegraphics[width=0.22\textwidth]{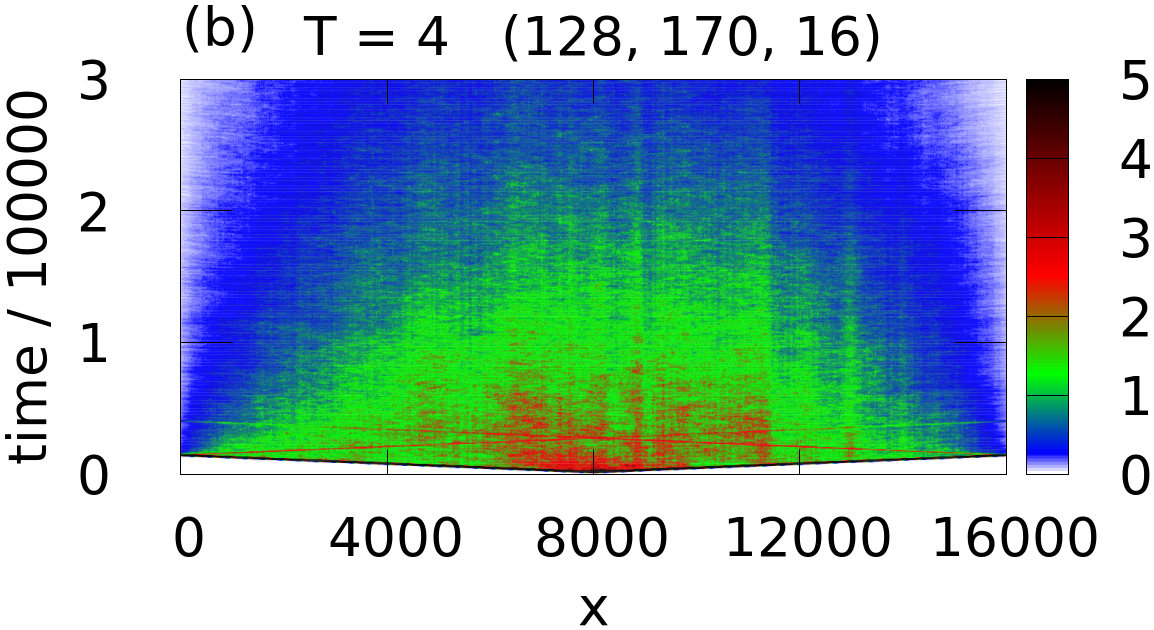}

~~\\
\includegraphics[width=0.22\textwidth]{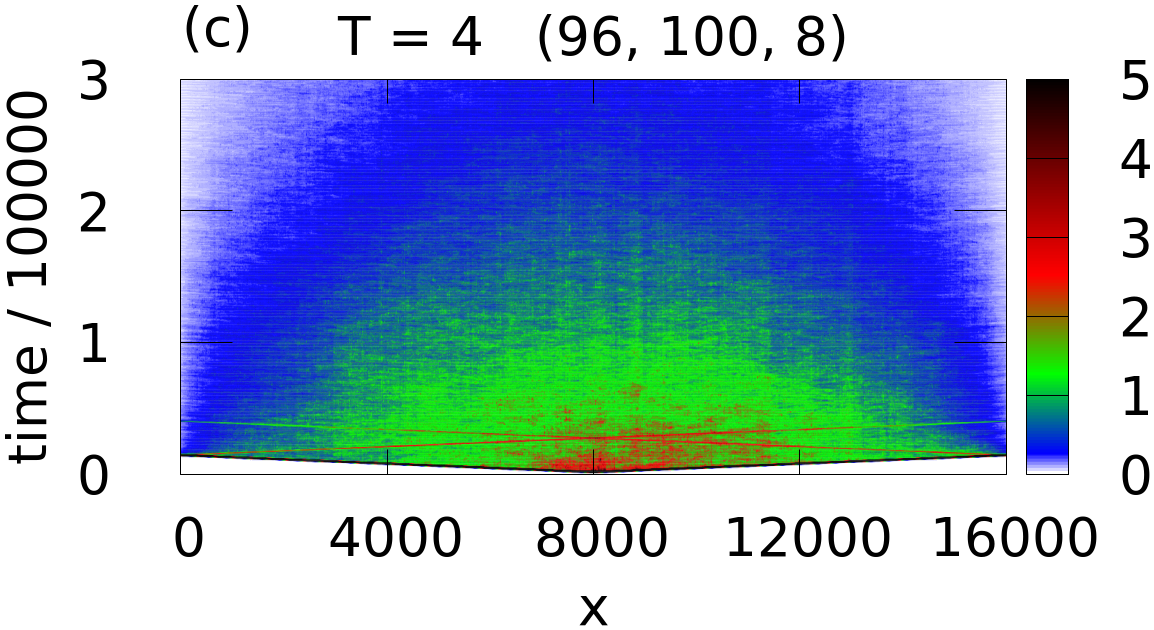}
\includegraphics[width=0.22\textwidth]{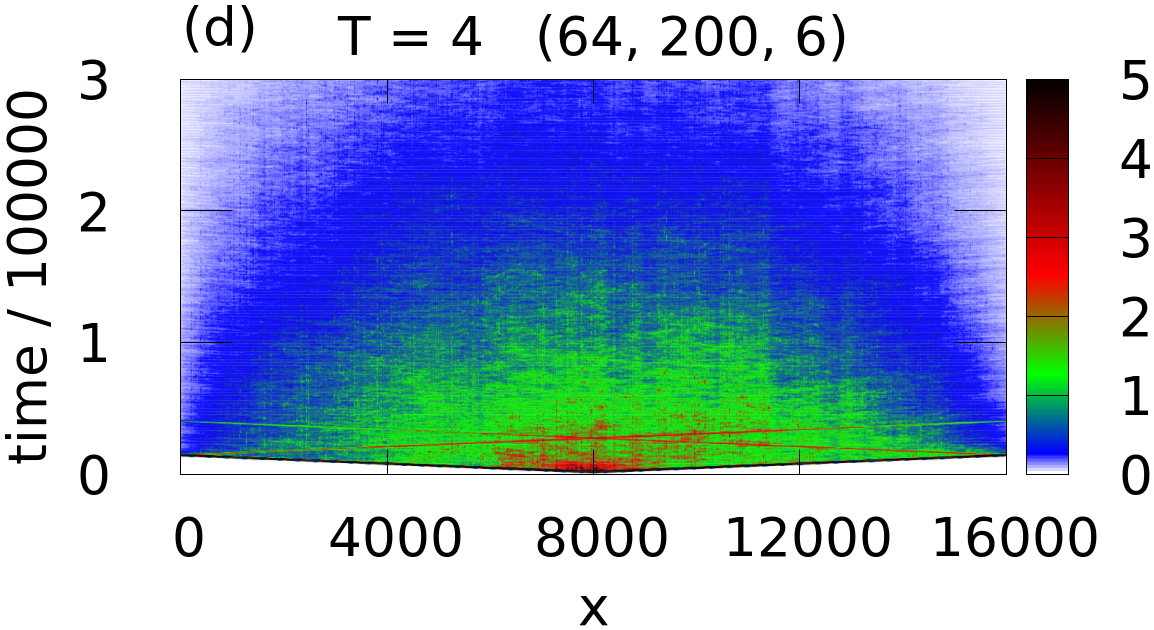}
\end{center}
\caption{%
	(Color online)
Normalized energy ${\cal E}(x,t)$ for
four system with different values of $(d, l_c, h)$ but similar value of  $z =l_c^{1/2}d^{3/2}/h\approx 1200$.
Period of the source $T=4$. Similarity of transport regimes in all four samples is supported also by calculation of $r^2$ shown in Fig. \ref{r2}(a).
}
\label{z1200}
\end{figure}

\begin{figure}[t]
\begin{center}
\includegraphics[width=0.45\textwidth]{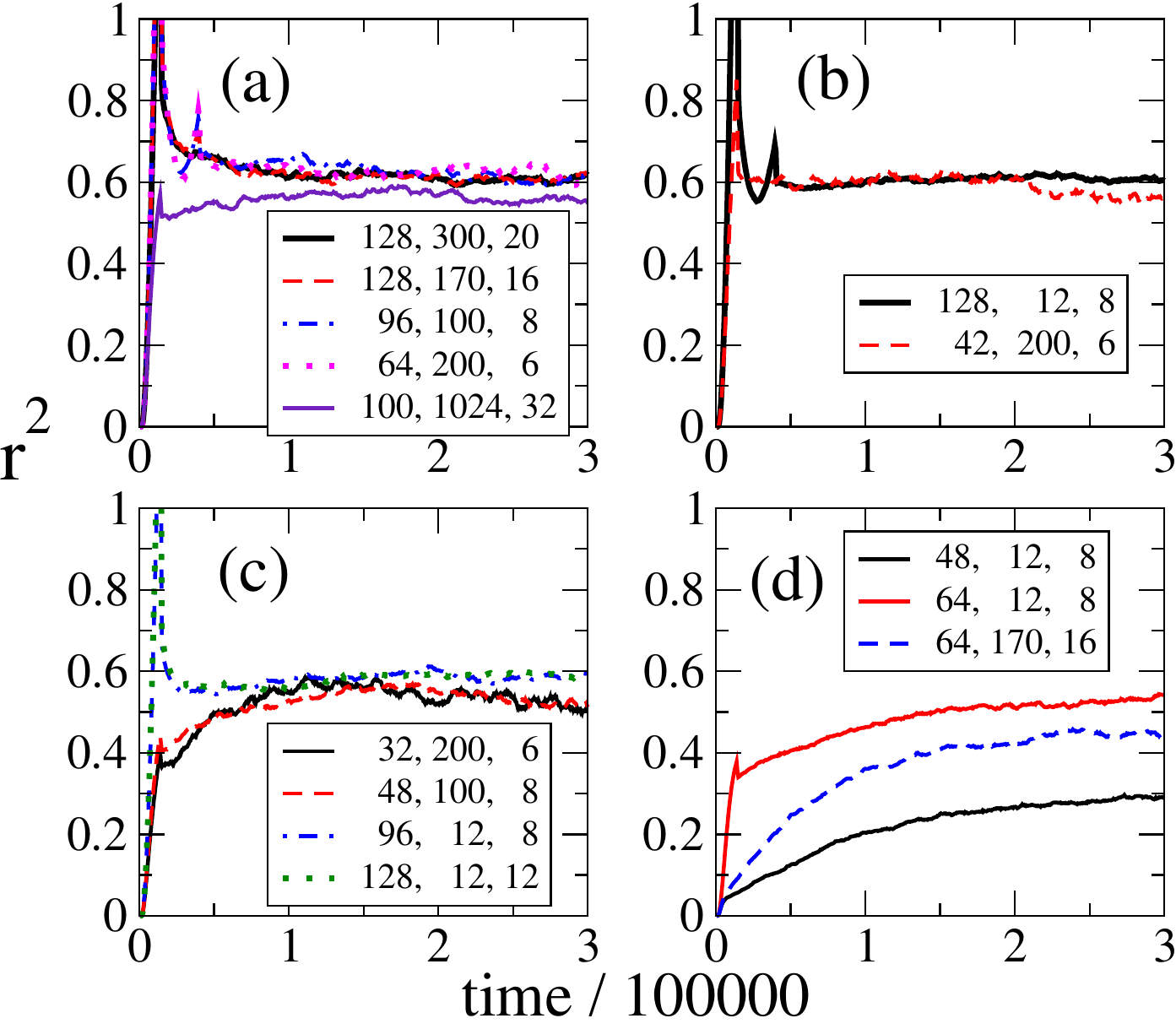}
\end{center}
\caption{%
(Color online)
	$r^2(t)$ for various systems illustrating universality. Period of the source $T=4$ for all samples.\cite{comment} 
Data confirm that the limiting value of $r_s^2 = \lim_{t\to\infty} r^2$ depends only on single parameter $z$ and decreases when $z$ decreases. 
The universality is clearly visible in (a) where $r^2$ for four different systems with $z\approx 1200$ converge to the same limiting value. 
Similarly, in (b), $z\approx 630$ and in (c) $z\approx 420$. Panel (d)
show data for systems with much  smaller values of $z$ which exhibit the phonon localization
 (dashed line corresponds to system where bottleneck effect dominates to make transmission more difficult).
}
\label{r2}
\end{figure}

 First, in order to verify the hypothesis of universality proposed in [\onlinecite{mm}], namely that the propagation of phonons with a given frequency 
depends only on $z=l^{1/2}_cd^{3/2}/h$, not independently on all the different length scales,
we simulated large number of various samples that differ in $d$, $l_c$ and $h$ and found that the transport of energy in various samples is similar if theses samples possess similar values of the parameter $z$. 
As an example, we show in Fig. \ref{z1200} the time and space evolution of the normalized energy ${\cal E}(x,t)$ for four disordered systems with  different
width and disorder but similar values of $z\sim 1200$.
More quantitatively, we plot in Fig. \ref{r2} the quantity $r^2(t)$ for various disordered samples. Our data show that
samples with similar value of $z$ exhibit similar time evolution  and similar saturation values  $r_s^2$. Also,  $r_s^2$ increases
when $z$ increases. This is consistent with the proposal that for a given frequency, localization appears in samples with  smaller values of $z$. 

\begin{figure}[t]
\begin{center}
\includegraphics[width=0.22\textwidth]{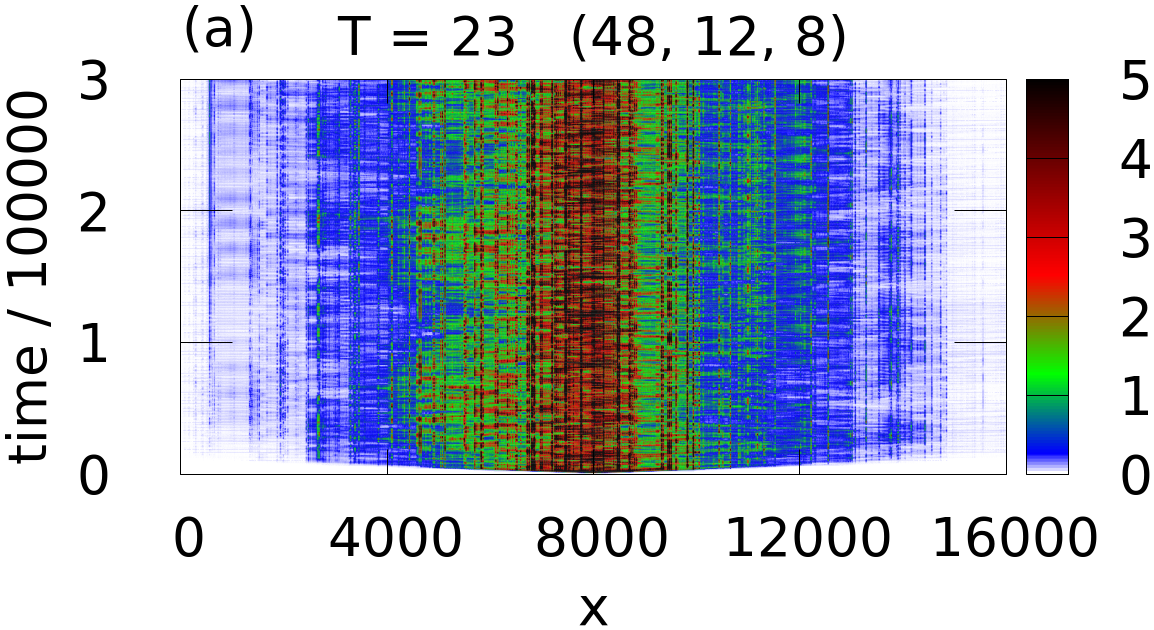}
\includegraphics[width=0.22\textwidth]{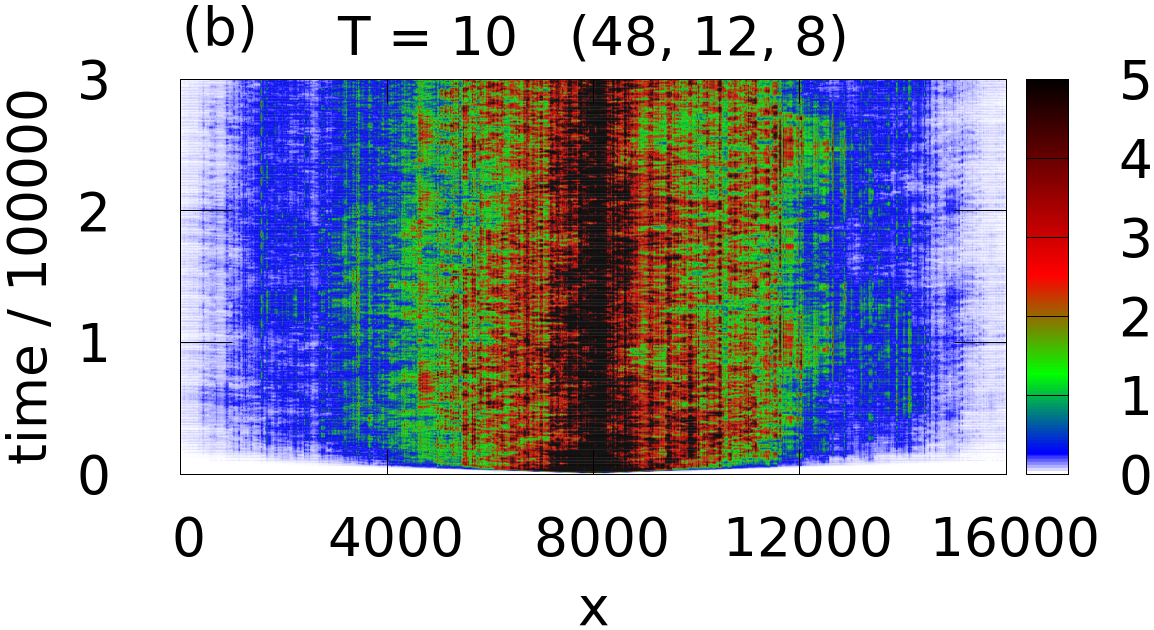}

~~\\
\includegraphics[width=0.22\textwidth]{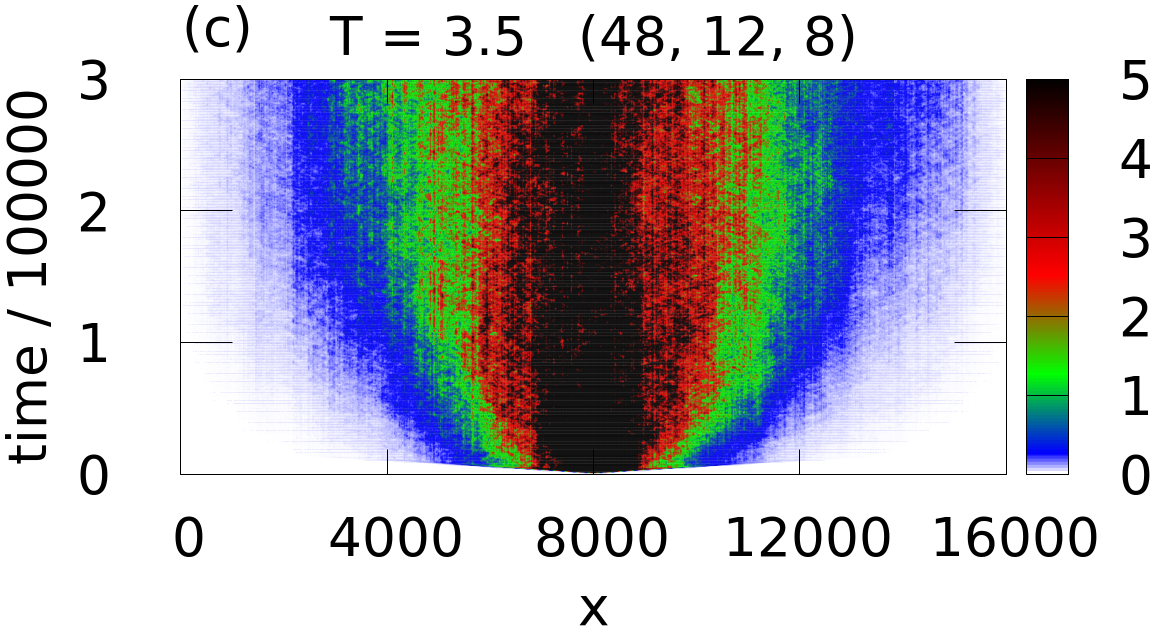}
\includegraphics[width=0.22\textwidth]{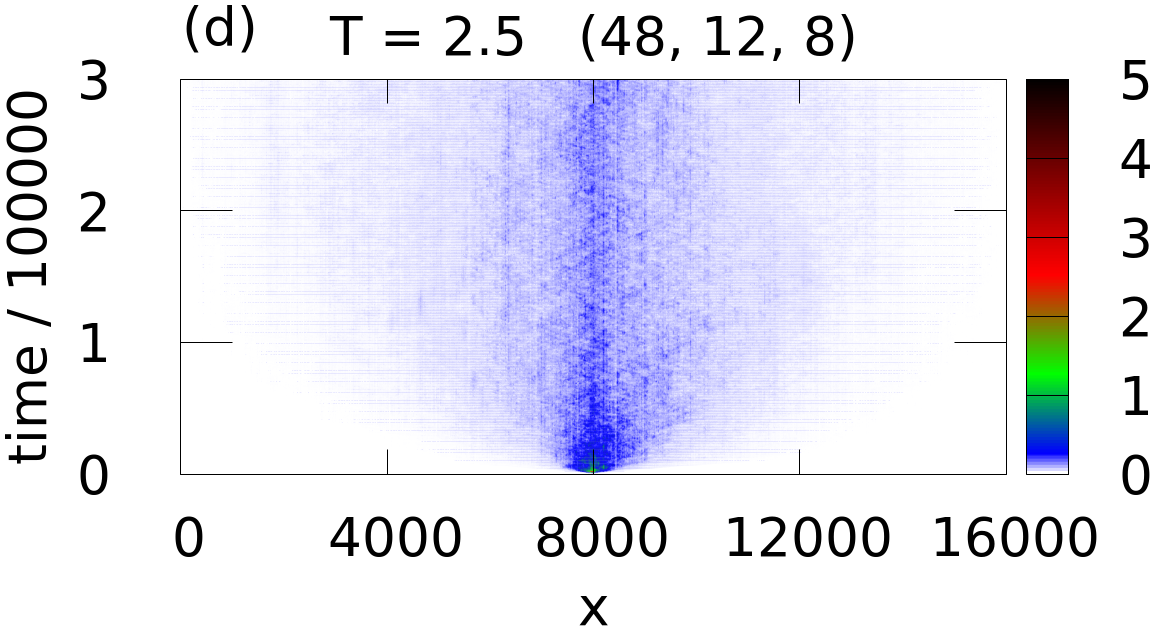}

~~\\
\includegraphics[width=0.22\textwidth]{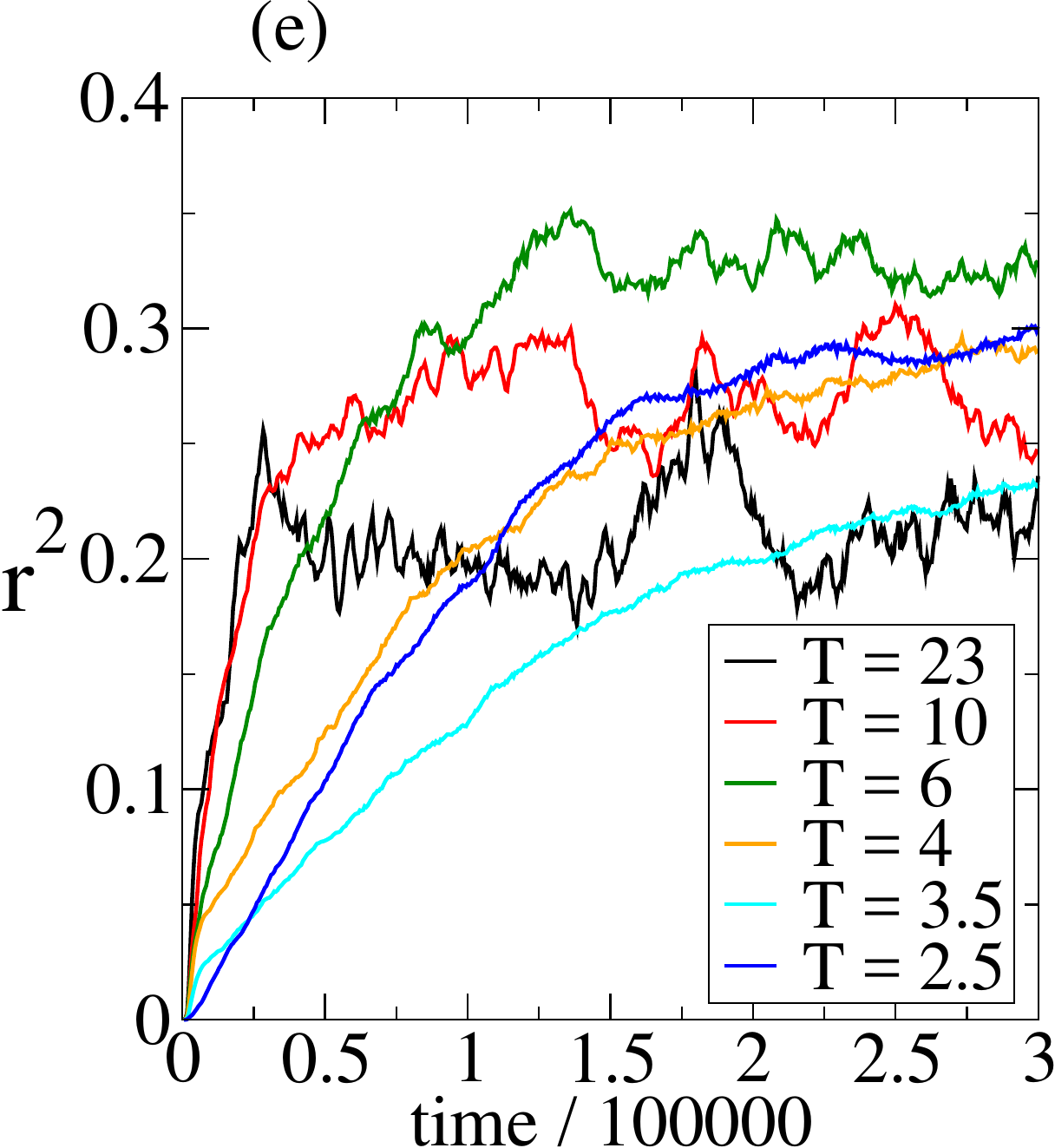}
\includegraphics[width=0.23\textwidth]{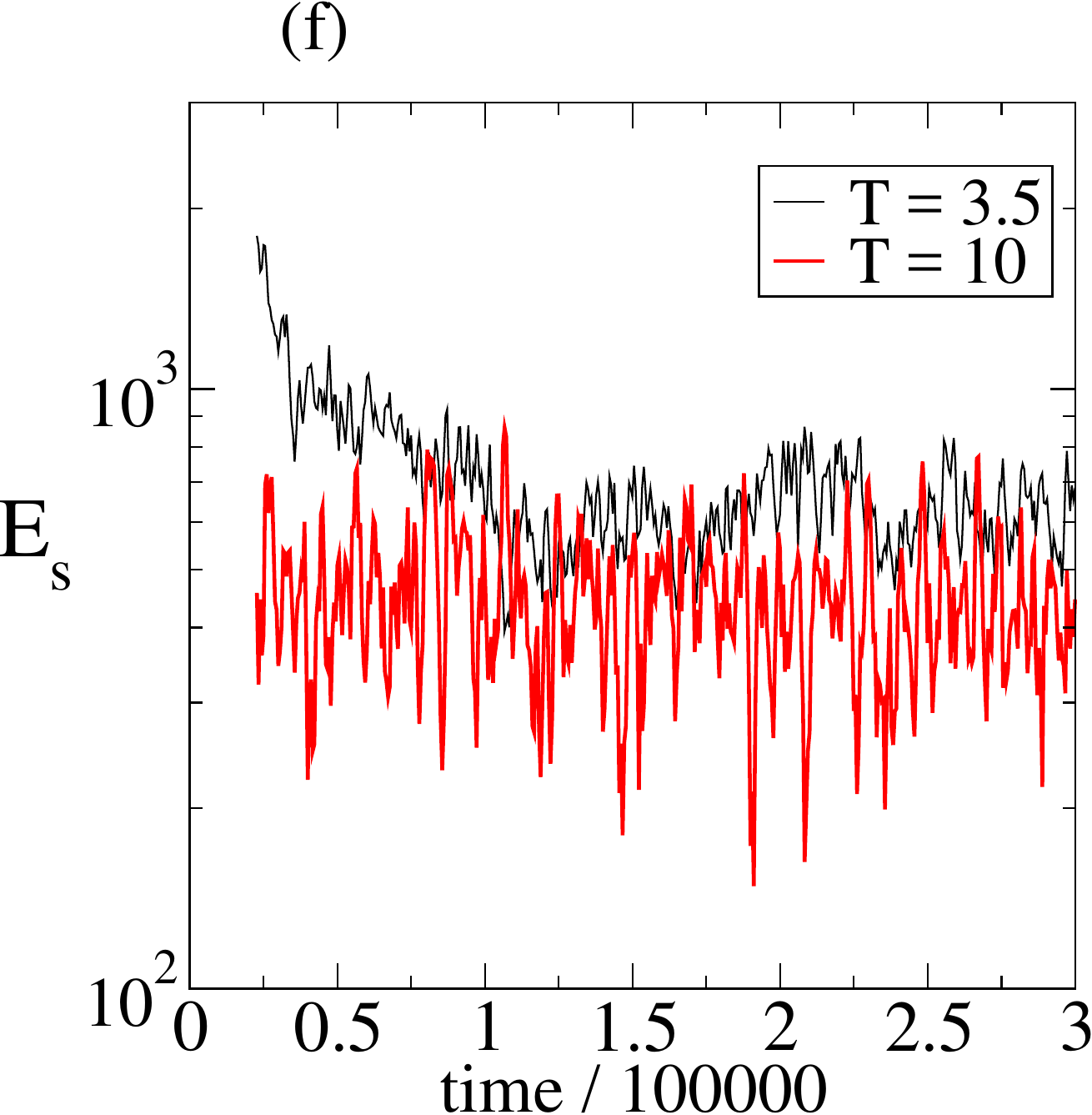}
\end{center}
\caption{%
	(Color online)
Localization of phonons in
strongly disordered system 
$48\times 16000$, $l_c=12$ and $h=8$   ($z=144$).
Panels (a-d)  show the normalized energy ${\cal E}(x,t)$  for periods $T=23, 10, 3.5$   and $2.5$.
(e)  time evolution of $r^2$,  (f)   the energy $E_s$ in the central region for periods $T=10$ and $T=3.5$.
}
\label{loc-48}
\end{figure}

For a small value of $z = 144$, we show 
in Fig. \ref{loc-48} the energy profile for the sample
$d=48, l_c=12, h=8$  for various periods $T$ of the source. 
Presented results indicate that phonons are localized for any period. For large $T=23$ (small frequency, left upper panel), the transport is 
suppressed  due to the bottleneck effects. This can be concluded from the long vertical lines which indicate strong reflection of waves at a given position.
For higher frequencies, localization appears due to  large number  of resonances.
The localization of phonons is confirmed also by the time evolution of the displacement   $r^2(t)$ which converges to rather small values for any period $T$ and 
by  time evolution of the energy $E_s$ in the central region of the sample which clearly does not decrease with time.

\begin{figure}[t]
\begin{center}
\includegraphics[width=0.30\textwidth]{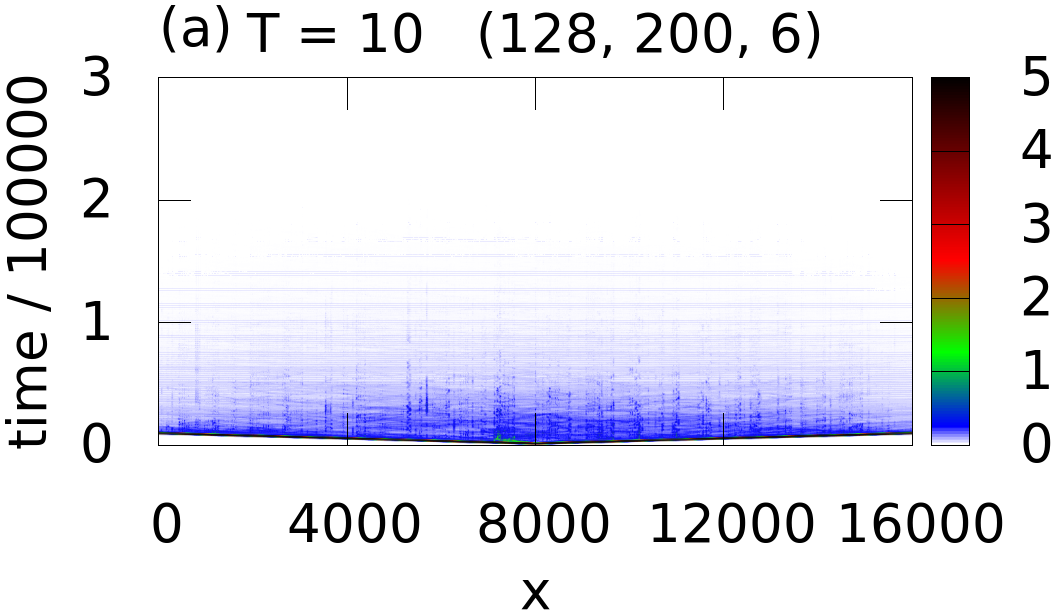}
\includegraphics[width=0.16\textwidth]{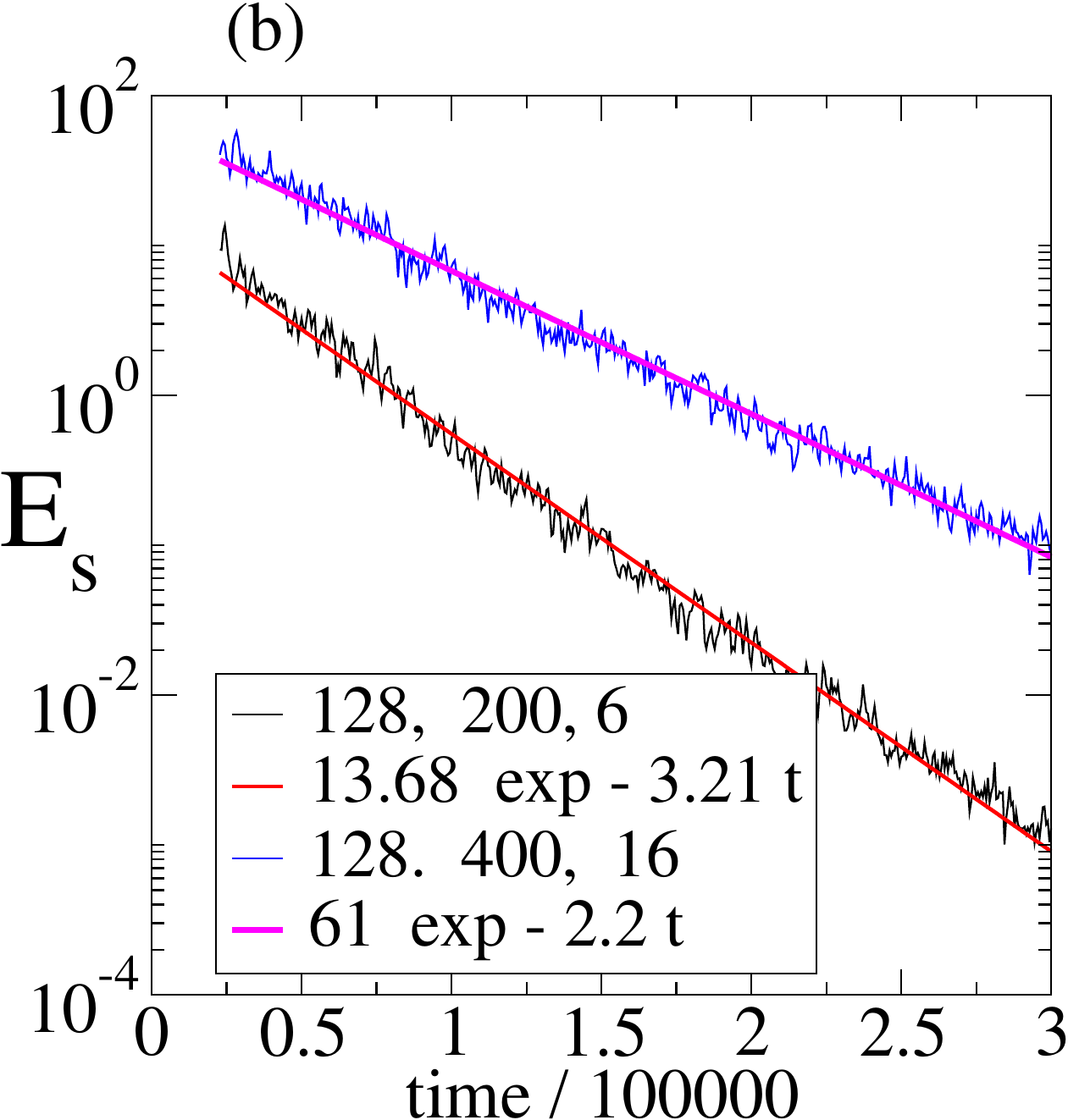}
\end{center}
\caption{%
	(Color online)
(a) normalized energy  ${\cal E}(x,t)$ for the system with very large value  $z = 3413$. Period of the source $T=10$.
(b) Exponential decrease of the energy from the central region.
}
\label{z3413}
\end{figure}

In the  opposite limit, weakly disordered samples with large values of $z$ exhibit fast decrease of the energy of the system.
An example is   shown in Fig. \ref{z3413}. 

\begin{figure}[t!]
\begin{center}
\includegraphics[width=0.23\textwidth]{mm-fig6a.png}
\includegraphics[width=0.23\textwidth]{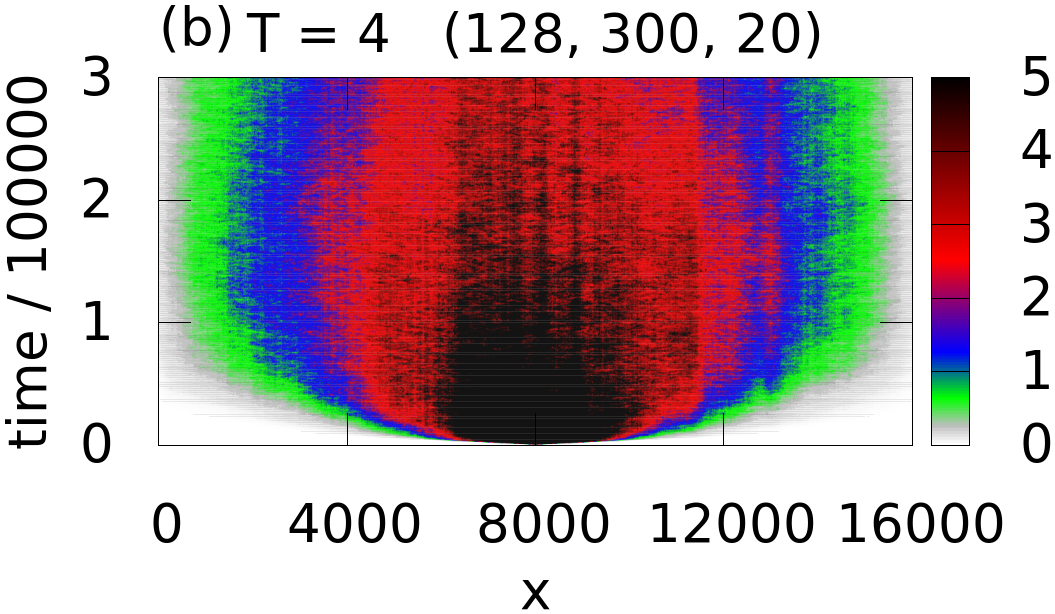}
\includegraphics[width=0.21\textwidth]{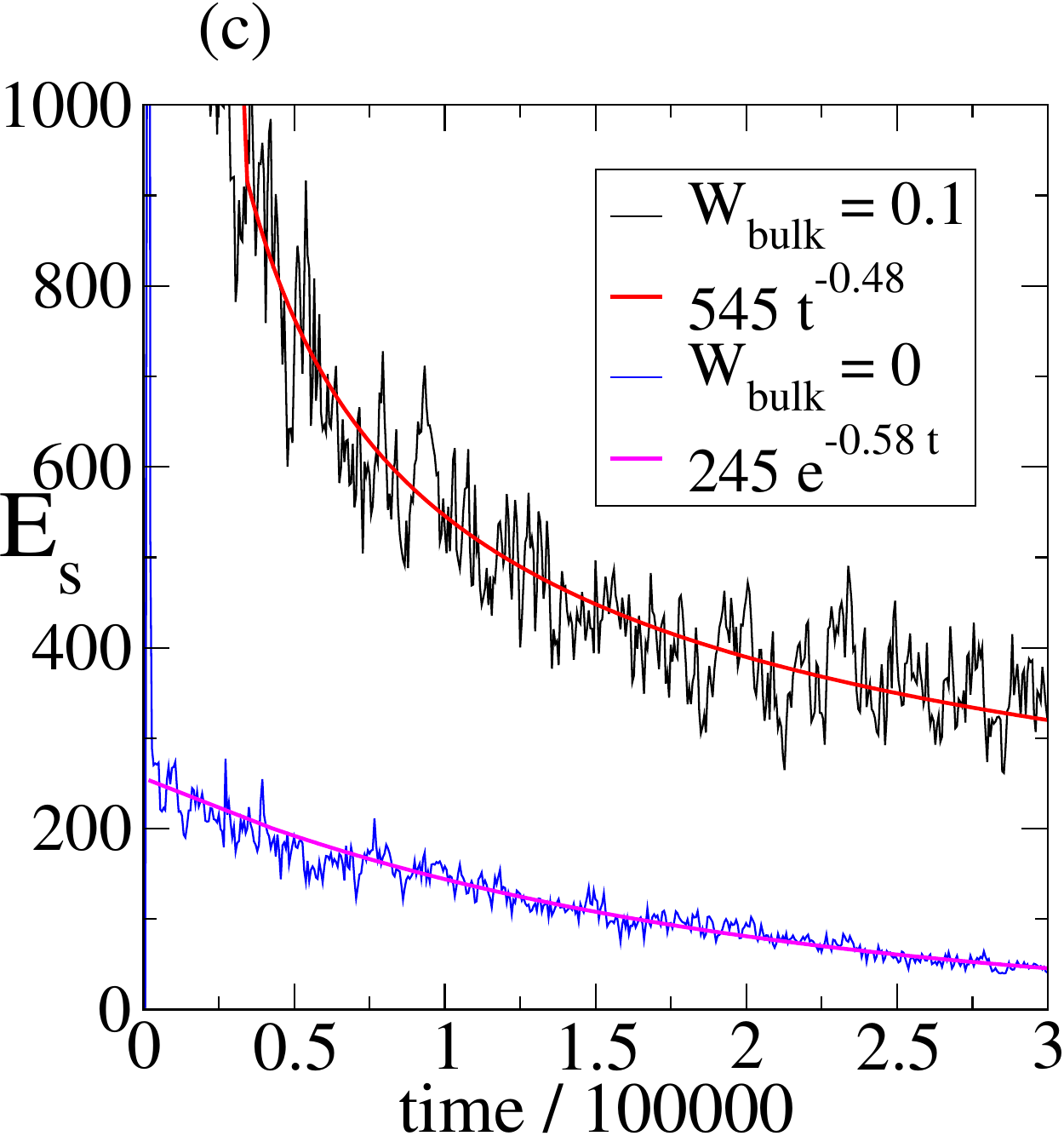}
~~
\includegraphics[width=0.21\textwidth]{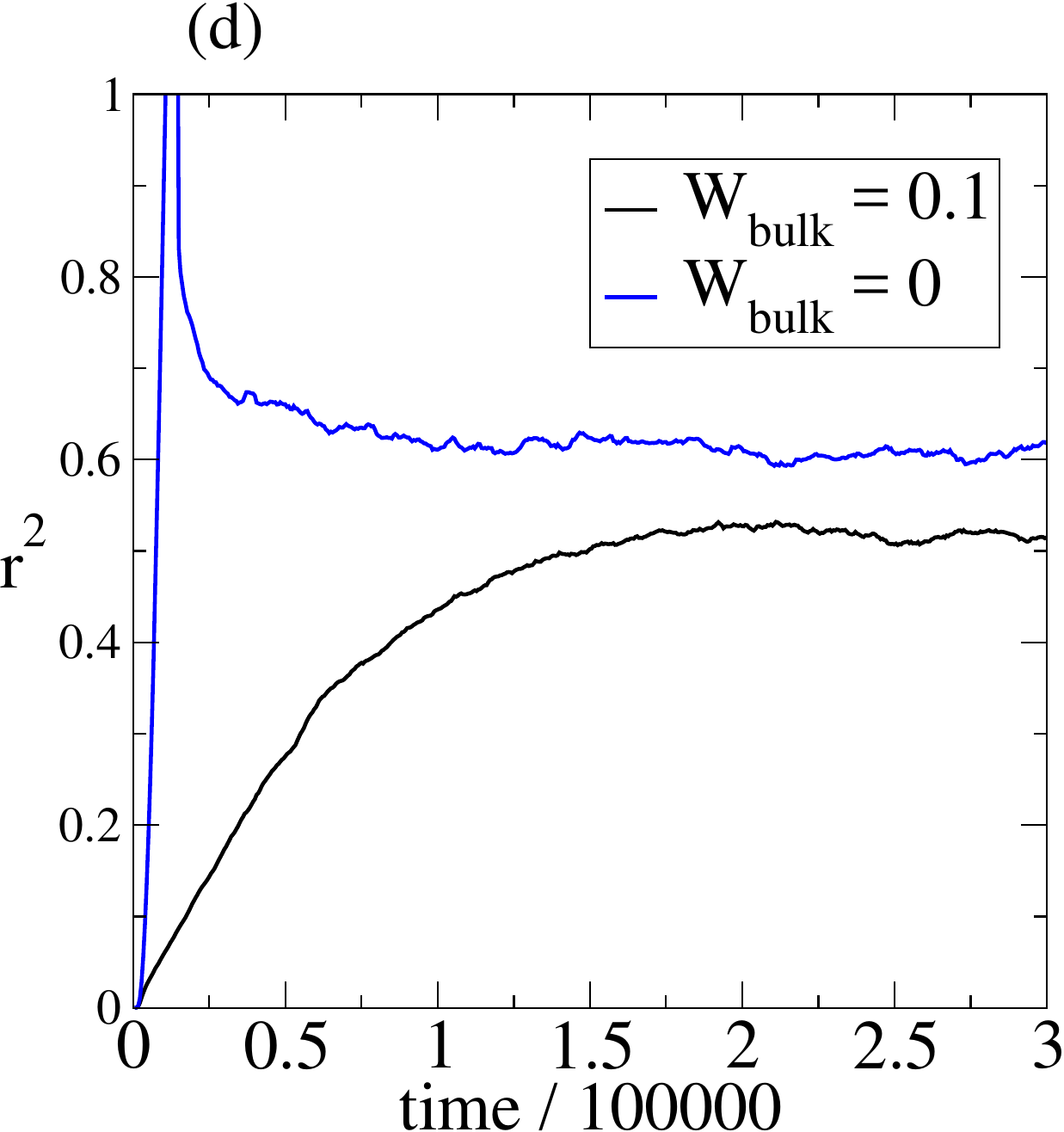}
\end{center}
\caption{%
	(Color online)
Comparison of sample without and with  bulk disorder $W_{\rm bulk}= 0.1$. 
Normalized energy ${\cal E}(x,t)$ for (a) $W_{\rm bulk} = 0$ and (b) $W_{\rm bulk} = 0.1$.
(c) Bulk disorder changes the time dependence of the energy $E_s(t)$  from exponential decrease to the  power-law. 
This suggests that diffusion is apparently more robust when bulk disorder is present. 
(d) $r^2$ for system with and without bulk disorder.
}
\label{bulk}
\end{figure}

Thus, as shown in Figs. \ref{loc-48} and \ref{z3413}, the time evolution of the energy $E_s(t)$ as well as the mean displacement $r^2(t)$ are  good quantitative measures of localization. 
In the  localized regime, the energy does not propagate away from the region of the source. 
In the highly delocalized regime, $E_s(t)$ 
decreases exponentially. This
exponential decrease, $E_s(t)\sim e^{-\alpha t}$,  is observed in various systems with large $z$, and the
exponent  $\alpha$ decreases when $z$ decreases. For smaller values of $\alpha$, it is difficult to distinguish between the 
exponential and a power-law behavior. 
We expect that adding a weak bulk disorder could make a power-law regime more robust, indicating a diffusive-like regime. Indeed, bulk disorder is always present in real samples, and it can be introduced easily by considering a small amount of different atoms 
(say, with mass $M=1.1M_0$). As shown in Fig. \ref{bulk}, this indeed generates a clear
power-law  decrease of the energy $E_s(t)$. 

\section{Summary and discussion}

\begin{figure}[t]
\begin{center}
\includegraphics[width=0.23\textwidth]{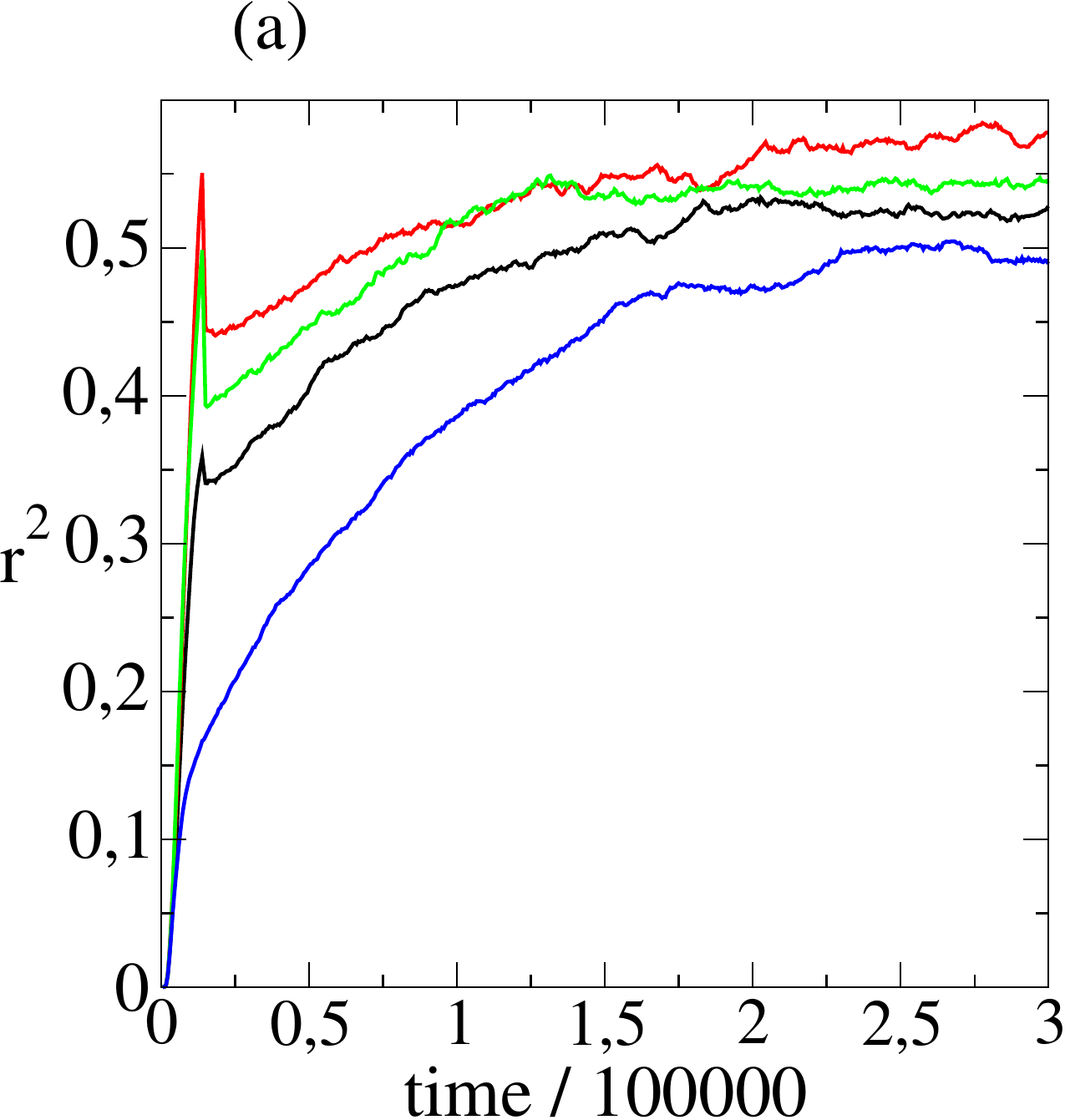}
~~
\includegraphics[width=0.23\textwidth]{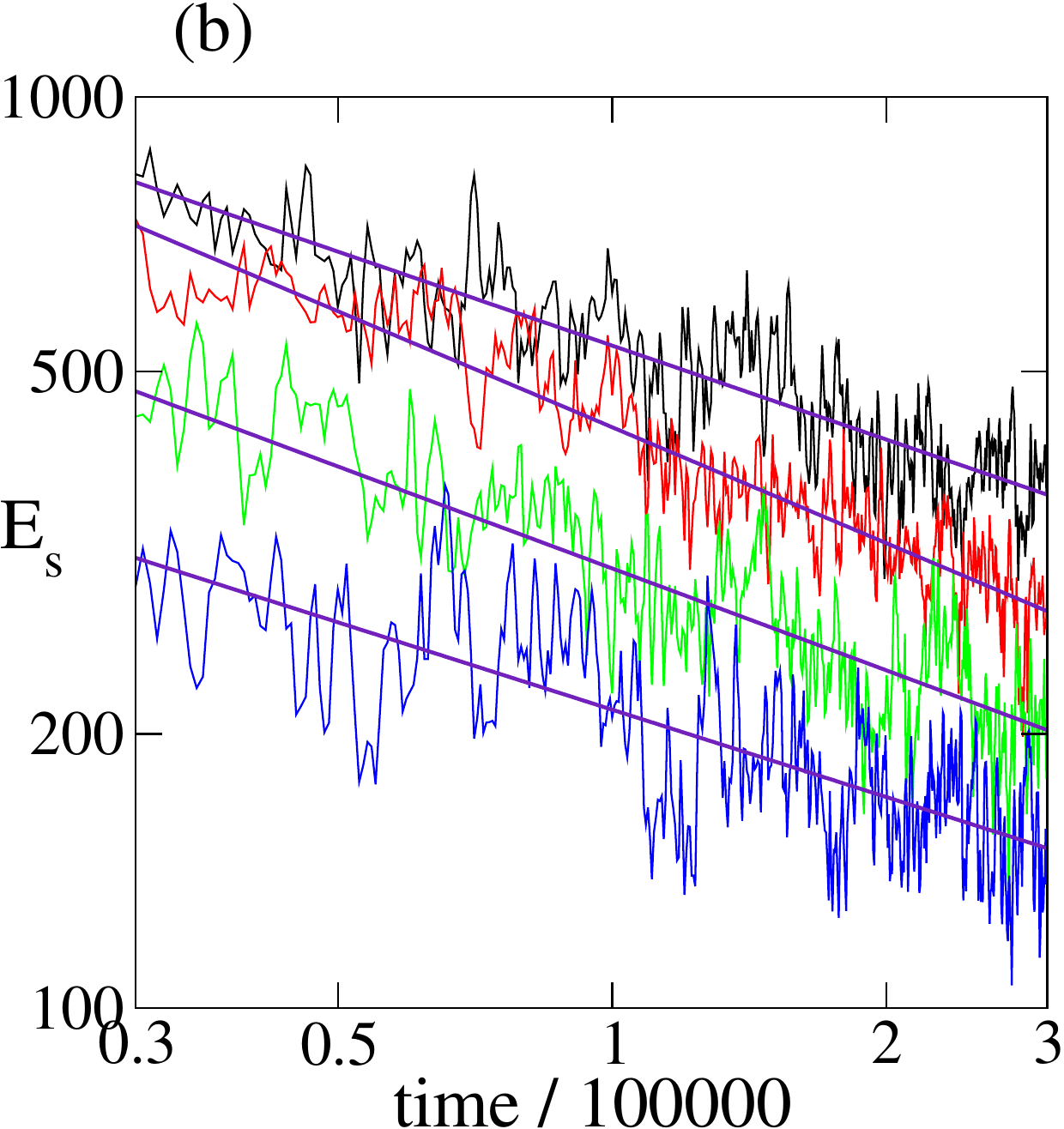}
\end{center}
\caption{%
	(Color online)
	(a) $r^2$ calculated for four different samples which differ by realizations of surface disorder. 
The size is $128\times 16000$, $l_c=200$ and $h=30$ ($z = 682$). Results give us some information about the statistical
properties of calculated quantities. 
(b)  Time evolution of the energy $E_s(t)$ in the center behaves as $E(t)\sim t^{-\alpha}$ with exponent $\alpha\sim 0.32 - 0.37$. 
}
\label{four}
\end{figure}
 
We argue that nanowires with rough surfaces are good candidates to observe localization of phonons experimentally. By studying numerically the space and time evolution of energy $E(x,t)$ across the length of a nanowire with surface disorder after a heat pulse is injected at some position, we show that it is possible to determine if localized phonons are excited at various positions that might affect the nature of energy transport in the wire. We confirm the universality proposed earlier \cite{mm} that for a wire of a given length, surface disorder is characterized by a single parameter $z=l_c^{1/2}d^{3/2}/h$ where $d$ is the diameter, $l_c$ is the correlation length and $h$ is the mean corrugation height,  smaller $z$ corresponding to larger disorder. 
While the values of the individual disorder parameters we choose for our simulation are not necessarily realistic, the universality allows us to consider values of the parameter $z$ that are experimentally accessible. 
For sufficiently small values of $z$, heat pulses (of frequencies within the band) injected into the wire excites long-lived resonances at various positions that depend on the frequency of the heat pulse. By considering the time evolution of the energy content $E_s(t)$ within a region near the pulse as well as the mean displacement $r^2(t)$ away from the pulse we show that energy remains localized for sufficiently large disorder.   
We show that as $z$ increases, localized phonons start to overlap and generate transport that is more diffusive like. Adding a weak bulk disorder makes this transport regime more robust.  

	Although the observation of individual localized phonons might be difficult in experiment, since it requires very narrow frequency pulse, we believe that the effect of localization is experimentally observable if the time evolution of the injected energy is measured along the sample. 

As we indicated in the text, heat transport experiments have observed very small thermal conductivity in surface disordered silicon nanowires \cite{hochbaum,lim}. While they are suggestive of the presence of localized phonons \cite{mm}, it is not a direct observation of phonon localization. Here we note that the values of the universal surface disorder parameter $z$ that corresponds to regimes where localized phonons are observed in our simulations are in the same range for the ELE silicon nanowires as reported in Lim et al \cite{lim}, where it varies from $z=390$ for a wire characterized by $(h,l_c,d)=(4.3,8.4,69.7\; {\rm nm})$  to $z=883$ for a wire with $(h,l_c,d)=(2.3,8.9,77.5 \; {\rm nm})$. Thus we propose that similar ELE nanowires should be good candidates for experimental studies of localization of phonons.  

We note that we have not done any ensemble averaging of $E(x,t)$ that requires considering a large number of samples with different realizations of disorder. Clearly, it would be useful to study such ensemble averaged quantities in order to obtain a more accurate time dependence of the energy. 
Figure \ref{four} shows one example for an intermediate $z=682$, where sample to sample fluctuations are already significant. Nevertheless, the qualitative features of the functions $r^2(t)$ or $E_s(t)$ are not sensitive to different realizations of disorder. For example the power-law decay of $E_s(t)$ for all four samples are similar, with exponent $\alpha$ that varies between $\alpha=0.32$ and $\alpha=0.37$. Moreover, our analysis of different combinations of length scales leading to similar values of $z$ can be thought of as one way of considering different realizations of disorder. Since it involves significant computational time, we argue that an extensive study of ensemble averaging is not necessary for our current limited purposes. It would of course be important if we need to obtain e.g. the exact power law in the diffusive transport regime or the values of $z$  where a crossover from the localized to diffusive regime might occur. This crossover is expected to depend on the frequency of the heat pulse, and again, a careful ensemble averaging would be needed to find any possible `phase boundary' in this disorder-frequency space. We leave these interesting questions for future considerations.

\section{acknowledgment}

We  acknowledge  financial  support  from  the  Slovak  Re-
search and Development Agency under Contracts No.  APVV-16-0372 and  APVV-14-0605  
and  from  the  agency  VEGA  under  Contract  No.
1/0108/17. K.A.M. acknowledges stimulating discussions with A. Majumdar and R. Chen.


\begin{thebibliography}{99}
\bibitem{comment} Sharp maximum observed for short time is due to ballistic phonons excited by source at the beginning of simulations which transmit a big portion to the energy.
\end{thebibliography}
\end{document}